 \documentclass[10pt,epsf]{article}
\usepackage{epsfig}
\usepackage{color}
\font\tenbf=cmbx9

\font\fourbf=cmbx12

\font\tenrm=cmr9

\font\tenit=cmti9
\font\tenitr=cmr9
\font\sc=cmr12
\font\tensc=cmr9

\font\lrm=cmr9

\newcommand{\lle}{\mbox{$\langle$}}
\newcommand{\rle}{\mbox{$\rangle$}}

\newcommand{\bfsigma}{\mbox{\boldmath$\sigma$}}

\newcommand{\bfsi}{\mbox{\boldmath$\sigma$}}
\newcommand{\bfep}{\mbox{\boldmath$\varepsilon$}}

\newcommand{\bfze}{\mbox{\boldmath$\zeta$}}

\newcommand{\bfde}{\mbox{\boldmath$\delta$}}

\newcommand{\bfcK}{\mbox{\boldmath$\cal K$}}
\newcommand{\bfcL}{\mbox{\boldmath$\cal L$}}

\newcommand{\bfg}{\mbox{\boldmath$\bf g$}}

\newcommand{\bfs}{\mbox{\boldmath$\bf s$}}

\newcommand{\bfx}{\mbox{\boldmath$\bf x$}}
\newcommand{\bfy}{\mbox{\boldmath$\bf y$}}
\newcommand{\bfz}{\mbox{\boldmath$\bf z$}}

\newcommand{\bfB}{\mbox{\boldmath$\bf B$}}
\newcommand{\bfC}{\mbox{\boldmath$\bf C$}}
\newcommand{\bfD}{\mbox{\boldmath$\bf D$}}
\newcommand{\bfE}{\mbox{\boldmath$\bf E$}}

\newcommand{\bfH}{\mbox{\boldmath$\bf H$}}
\newcommand{\bfI}{\mbox{\boldmath$\bf I$}}

\newcommand{\bfL}{\mbox{\boldmath$\bf L$}}
\newcommand{\bfN}{\mbox{\boldmath$\bf N$}}

\newcommand{\bfQ}{\mbox{\boldmath$\bf Q$}}

\newcommand{\bfY}{\mbox{\boldmath$\bf Y$}}
\newcommand{\bfZ}{\mbox{\boldmath$\bf Z$}}

\newcommand{\bfF}{\mbox{\boldmath$\bf F$}}
\newcommand{\bfR}{\mbox{\boldmath$\bf R$}}
\newcommand{\bfK}{\mbox{\boldmath$\bf K$}}
\newcommand{\bfM}{\mbox{\boldmath$\bf M$}}
\newcommand{\bfT}{\mbox{\boldmath$\bf T$}}
\newcommand{\bfS}{\mbox{\boldmath$\bf S$}}

\newcommand{\bfcD}{\mbox{\boldmath$\cal D$}}

\newcommand{\bfcT}{\mbox{\boldmath$\cal T$}}

\newcommand{\bfcR}{\mbox{\boldmath$\cal R$}}
\newcommand{\bfcB}{\mbox{\boldmath$\cal B$}}

\newcommand{\bfbe}{\mbox{\boldmath$\beta$}}

\newcommand{\bfGa}{\mbox{\boldmath$\Gamma$}}
\newcommand{\bfet}{\mbox{\boldmath$\eta$}}
\newcommand{\bfbeta}{\mbox{\boldmath$\beta$}}

\newcommand{\bfeta}{\mbox{\boldmath$\eta$}}

\newcommand{\BB}{\begin{equation}}
\newcommand{\EE}{\end{equation}}

\newcommand{\BBEQ}{\begin{eqnarray}}
\newcommand{\EEEQ}{\end{eqnarray}}
{\tenit V. A. Buryachenko
\hfil}
{\tenit \hfil On the
thermo-elastostatics of heterogeneous materials. II}

\begin{document}
%\noindent{\lrm  Acta Mech (2009)000-000, {\tenbf Manuscript N09-}}

\vspace{28pt}
\noindent {\bf Valeriy  A. Buryachenko}\footnote{\tenrm
%V. A. Buryachenko\\
Department of Structural Engineering, University of Cagliari, 09124 Cagliari, Italy;
%University of Dayton Research Institute, Dayton, OH 45469-0168, USA;\\
E-mail: Buryach@aol.com}

\vspace{15pt}
%{\baselineskip=16pt

\noindent{\fourbf On
%some background of
the thermo-elastostatics of heterogeneous materials}

\vspace{6pt}

\noindent{\fourbf  II. Analyze and generalization of some basic hypotheses and propositions}

%\vspace{6pt}

%\noindent{\fourbf  and propositions}

\vspace{60pt}
%\noindent{\tenrm Received: July, 2009}

%\vspace{12pt}

\centerline{\vrule height 0.003 in width 16.0cm}
 \noindent {\baselineskip=9pt
{\tenbf Abstract} {\tenrm
One considers linearly thermoelastic  composite media, which consist of a homogeneous matrix containing a statistically homogeneous  random set of ellipsoidal
uncoated or coated inclusions. Effective properties (such as  compliance and thermal
expansion) as well as the first  statistical moments of stresses in the phases are estimated for the general case of nonhomogeneity of the thermoelastic inclusion properties.
At first, one shortly reproduces both the basic assumptions and propositions of micromechanics used in most popular methods, namely: effective field hypothesis, quasi-crystallite approximation, and the hypothesis of ``ellipsoidal symmetry".
The explicit new representations of the effective thermoelastic properties and
stress concentration factor are expressed through
 some building blocks described by numerical solutions
for both the one and two inclusions inside the infinite medium subjected to both the homogeneous and inhomogeneous remote loading. The method uses as a background the new general integral equation proposed in the accompanied paper and makes it possible to abandon the basic concepts of micromechanics mentioned above. The results of this abandonment are quantitatively estimated for some modeled composite reinforced by aligned continously inhomogeneous fibers. Some new effects are detected that are impossible in the framework of a classical background of micromechanics.
\par}}
\vspace{-5pt}

\noindent
{\lrm Keywords: A. microstructures, B. inhomogeneous material, B.
 elastic material.}

%\smallskip
 \centerline{\vrule height 0.003 in width 16.0cm}

\bigskip
\noindent {\bf 1. Introduction}
\medskip

The prediction of the behavior of composite materials
in terms of the mechanical properties of constituents and their
microstructure is a central problem of micromechanics,
which is evidently reduced to the estimation of stress
fields in the constituents. Appropriate, but by no means
exhaustive, references for the estimation of effective
elastic moduli of statistically homogeneous media are
provided by the reviews [1-7].
% of Willis (1981),
% && , 1982, 1983),
% Mura (1987), Nemat-Nasser and Hori  (1993), Torquato (2002), Milton (2002),
% Buryachenko (2007), Kanaun and Levin (2008).
It appears today that variants
of the effective medium method by Kr\"oner [8]  %, 1958;
and by Hill [9],  % , 1965)
and the mean field method [10], [11] %(Mori and Tanaka, 1973; Benveniste, 1987)
are the most popular and widely used methods.
Recently a new method has become known, namely the multiparticle effective field method (MEFM) was put forward and developed by the author
(see for references Buryachenko [6]). %, 2007).
The MEFM is based on the theory of functions of random variables and Green's functions. Within this method one constructs a hierarchy of statistical moment equations for conditional averages of the stresses in the inclusions. The hierarchy is then cut by introducing the notion of an effective field. This way the interaction of different inclusions is taken into account. Thus, the MEFM does not make use of a
number of hypotheses which form the basis of the traditional one-particle methods.

However, a diversity of micromechanical methods and their specific formulations astonish our imagination only at first glance. We will see that most popular methods are based just on a few basic concepts of micromechanics.
Effective field hypothesis is apparently the most fundamental, most prospective, and most exploited concept of micromechanics. This concept has directed a development of micromechanics over the last sixty years and made a contribution to their progress incompatible with any another concept.
The notion of an effective field in which each particle is
located is a basic concept of such powerful methods in micromechanics
as the methods of self-consistent fields and effective fields
(for references see [6], [7], [12].
%Morse and Feshbach,  1953; Buryachenko, 2007; Kanaun and Levin, 2008).
The idea of this concept dates back to Mossotti [13] %(1850)
and Clausius (in the
dielectric context), Lorenz (in the refractivity context), and Maxwell (in the
conductivity context). Markov [14] % (1999)
and Scaife [15] %(1989)
presented comprehensive
reviews of the 150-year history of this concept accompanied by some famous formulae with extensive references. Mossotti [13] % (1850)
 (especially Clausius) pioneered the introduction of the effective field concept
as a local homogeneous field acting on the inclusions and differing from the applied macroscopic one. Among a few hypotheses used by Mossotti [13], %(1850),
the most important one was in fact the quasi-crystalline approximation proposed 100 years later by Lax [16]  %(1952)
in a modern concise form. The concept of the effective field in combination with
subsequent assumptions was introduced in a modern formalized form in the physics of multiple scattering of waves (see, e.g.,  [16-18]). % Lax, 1952; Foldy, 1945;; Chaban, 1965).
 Walpole [19] %(1966)
pioneered the application of the concept to the static of composites under the name uniform image field.  Effective field technique was intensively applied in micromechanics of random and periodic structure composites (for references see, e.g.,  [6], [7])
%Buryachenko, 2007; Kanaun and Levin, 2008)
as well in micromechanics of multiple interacting cracks under the name traction
% &&(for references see, e.g., Kachanov, 1993)
or pseudo-load  [20]. %(Hori and Nemat-Nasser, 1987).
Buryachenko and Rammerstorfer [21] %(2000)
has drawn the conclusion that the effective field concept is used (either explictly or implicitly) in most popular methods of micromechanics such as, e.g., the effective medium method and their modifications, differential scheme, Mori-Tanaka method, and, needless to say, the MEFM.

The idea of effective field and quasi-crystalline approximation were added by
the hypothesis of ``{\it ellipsoidal symmetry}" for the distribution of inclusions attributed to Willis [22]. %(1977).
As a tool for concrete applications of the concepts mentioned, the Eshelby [23] %(1957)
solution was used although the Eshelby's theorem has a fundamental conceptual sense
(it will be shown in the current paper) rather than only an analytical solution of some particular problem for the ellipsoidal homogeneous inclusion. All these concepts creating the framework and background of modern statistical analytical micromechanics were transformed by the use of both the additional assumptions and sophisticated analytical and numerical tools to a few particular methods. However, we will show in this paper that the effective field hypothesis (also called the hypothesis {\bf H1a} is a central one and other concepts play a satellite role providing the conditions for application of the effective field hypothesis. Moreover, we will show that all mentioned hypotheses are not really necessary and can be relaxed.

The outline of the study is as follow. In Section 2 we recall the basic concepts defining the background of micromechanics. The interconnection between the different concepts and their essence are established.  In Section 3 the auxiliary problem for one inclusion in the infinite matrix is presented for a general remote loading. The new general integral equation obtained in an accompanying paper by Buryachenko [24], %(2009),
henceforth referred to as (I), is presented in Section 4 through the operator forms of the particular solutions for both one and two interacting inclusions.
This equation is solved by the iteration method in the framework of the quasi-crystallite approximation but without basic hypotheses of classical micromechanics such as both the effective field hypothesis and ``{\it ellipsoidal symmetry}" assumption.
In Section 5 we qualitatively explain the advantages of the new approach with respect to the classic ones and demonstrate the corrections of popular propositions obtained in the framework of the old background of micromechanics.  Quantitative estimations of results of the abandonment of the central hypothesis {\bf H1a} are presented in Section 6.

\bigskip
\noindent {\bf 2. Preliminaries. Basic assumptions and propositions of micromechanics}
\smallskip

\noindent{\it 2.1 General integral representations and notations}
\smallskip

\setcounter{equation}{0}
\renewcommand{\theequation}{2.\arabic{equation}}
For the sake of brevity of the current presentation, the basic equations of thermoelasticity, the homogeneous boundary conditions (2.5I), statistical description of the composite microstructure, assumption, and notations exploited in the current paper are presented in the accompanied paper by  Buryachenko [24] %(2009)
and the interested reader is referred to this publication, henceforth referred to as (I).

In this section we will shortly reproduce both the basic assumptions and propositions of micromechanics in the form adopted for subsequent presentation. In most detail we will consider the mentioned concepts as applied to the MEFM based on some mathematical approximations for solving the infinite systems of integral equations involved, although
other methods exploiting these concepts will  also be discussed.

For simplicity, we will consider only statistically homogeneous media
(described, as a particular case,  in Section 2.2 in I) subjected to the homogeneous boundary conditions (2.5I). If elastic properties of the comparison medium and matrix coincide (3.20I) then the known general integral equation in terms of stresses  (see, e.g., [6],I) %Buryahenko, 2007)
\BBEQ
\bfsi ({\bf x})&=&
\langle \bfsi\rangle
+\int\! [{\bf \Gamma} (\bfx-\bfy) \bfeta({\bf y})
-\langle {\bf \Gamma} (\bfx-\bfy)\bfeta\rangle ({\bf y})] d{\bf y},\\%(2.1)
\bfsi ({\bf x})&=&\langle \bfsi\rangle +\int
 {\bf \Gamma} (\bfx-\bfy)[ \bfeta({\bf y})-\lle\bfeta\rle(\bfy)]
d\bfy, % \eqno (2.2)$$
\EEEQ
can be much easier to solve because the stress-strain fields  can be studied inside the inhomogeneities but not in the matrix; here $\bfet=\bfM_1\bfsi+\bfbe_1$ and $\bfGa$ are the strain polarization tensor (3.14I) and the Green stress tensor (3.5I), respectively. Buryachenko [6, I] proved that  for no {\it long-range} order assumed,
and for $\bfx\in w$ considered in Eqs. (2.1) and (2.2) and removed far enough from the boundary $\Gamma$ ($a\ll |\bfx-\bfy|,\ \forall \bfy\in \Gamma$),
the right-hand side absolutly convergent integrals in (2.1) and (2.2) do not depend
on the shape and  size of the domain $w$, and they can be replaced by the
integrals over the whole space $R^d$. With this assumption we hereafter
omit explicitly denoting $R^d$ as the integration domain in the equations.
The new exact equation (2.1) forming a new background of micromechanics yields the known approximate one (2.2) only at some additional assumptions [(3.23I) or (3.24I), see for details I].

The solution of Eqs. (2.1) and (2.2) by the use of different assumptions provides the estimations of both the effective compliance ${\bf M}^*$
 and the effective eigenstrains $ \bfbeta ^ *$
governed by the overall constitutive relation
\BB
\langle \bfep \rangle =\bfM^* \langle \bfsi \rangle +\bfbe^*
%(2.3)
\EE
and defined by general relations
\BBEQ
{\bf M}^*= \bfM^{(0)}+\langle \bfM_1\bfB^* \rangle,\\ %(2.4)
\bfbeta^*= \bfbe^{(0)}+
\langle \bfB^{*\top}\bfbeta_1 \rangle ,
%\eqno (2.5)$$
\EEEQ
where ${\bf B}^*={\bf B}^*({\bf x})$ ($\bfx\in v$) is a local stress concentration
tensor in the inhomogeneities obtained under pure mechanical loading ($\bfbe\equiv{\bf 0}$)
\BB
\bfsigma({\bf x})={\bf B}^*({\bf x})\lle\bfsigma\rle\quad {\rm for}\quad
{\bf x}\in v.
%\eqno (2.6)$$
\EE

Analysis of Eqs. (2.1)-(2.6) leads to the universally accepted

\noindent {\bf Proposition 1.} {\it For statistically homogeneous media subjected to the homogeneous boundary conditions, linear elastic effective properties $\bfM^*$ and $\bfbe^*$ depend only on stress distributions inside the inhomogeneities $v$ but not inside the matrix $v^{(0)}$.}

Let the inclusions  $v_1,\ldots,v_n$ be fixed
and we define two sorts of effective fields
$\overline{\bfsi}_i(\bfx)$ and  $\widetilde{\bfsigma}
_{1,\ldots,n}(\bfx)\quad (i=1,\ldots,n;\ {\bf x}\in v_1,\ldots,v_n)$
by the use of the rearrangement of Eq. (2.1) [or (2.2)] in the following
form (see for the earliest references of related manipulations [6]):
% Buryachenko, 2007):
\BBEQ
\bfsi(\bfx)\!\!\!&=&\!\!\!\overline{\bfsi}_i(\bfx)+\int{\bf \Gamma}(\bfx-\bfy)V_i(\bfy)
\bfet(\bfy)d\bfy,\nonumber\\
\overline{\bfsi}_i(\bfx)\!\!\!&=&\!\!\!\widetilde {\bfsigma}
_{1,\ldots,n}(\bfx)+\sum_{j\neq i}\int
{\bf \Gamma (x-y)}V_j({\bf y})\bfet({\bf y})d{\bf y},\nonumber\\
\widetilde {\bfsigma}_{1,\ldots,n}(\bfx)
\!\!\!\!&=&\!\!\!\!\langle \bfsigma\rangle ({\bf x})
%\nonumber\\
+\int \Big\{{\bf \Gamma (x-y)}  \bfet(\bfy)
V(\bfy\vert;v_1,{\bf x}_1;\ldots;v_n,{\bf x}_n)
%\nonumber\\
%\!\!\!\!&-&\!\!\!\!
-\langle \bfGa(\bfx-\bfy)\bfet\rangle ({\bf y})\Big\} d{\bf y},
%(2.7)
\EEEQ
 for  ${\bf x}\in v_i,\ i=1,2,\ldots,n$; here
$V(\bfy\vert; v_1,{\bf x}_1;\ldots ;v_n,{\bf x}_n)$ is
   a random indicator function of inclusions $\bfx\in v$ under
   the condition that ${\bf x}_i\neq\bfx_j$ if $i\neq j$ ($i,j=1,\ldots, n)$.
Then, considering  some conditional statistical averages of
the general integral equation (2.1) leads
      to   an infinite system of {\it new integral equations} $(n=1,2,\ldots)$
\BBEQ
\!\!\!\!&&\!\!\!\!
\langle \bfsigma \vert v_1,{\bf x}_1;\ldots ;v_n,{\bf x}_n
\rangle ({\bf x})
 -\sum^n_{i=1}\int{\bf \Gamma (x-y)}\langle V_i({\bf y})
\bfet \vert v_1,{\bf x}_1;\ldots;v_n,{\bf x}_n\rangle ({\bf y})
d{\bf y}\nonumber\\
\!\!\!\!&=&\!\!\!\!\langle \bfsigma\rangle ({\bf x})
+\int \bigl\lbrace{{\bf \Gamma (x-y)}\langle \bfet
\vert;v_1,{\bf x}_1;\ldots;v_n,{\bf x}_n\rangle ({\bf y}) -
\langle \bfGa(\bfx-\bfy)\bfet\rangle ({\bf y})}\bigr\rbrace d{\bf y}.
%(2.8)
\EEEQ
Since ${\bf x}\in v_1,\ldots,v_n$
in the $n$-th line of the system can take the values of the
inclusions $v_1,\ldots,v_n$, the $n$-th line actually contains $n$ equations.
 The definitions of the effective fields $\overline {\bfsigma}_i(\bfx)$,
$\widetilde{\bfsigma}_{1,2,\ldots,n}(\bfx)$
as well as their statistical averages $\langle \overline {\bfsigma}_i\rangle(\bfx)$,
$\langle\widetilde{\bfsigma}_{1,2,\ldots,n}\rangle(\bfx)$
are  nothing more than
notation convenience for different terms of the infinite systems (2.7) and (2.8),
respectively.
The physical meaning of these fields and their graphic illustrations  are presented in Ref. [6].

\medskip
\noindent{\it 2.2 Approximate effective field hypothesis}
\medskip

In order  to simplify the exact system (2.8) we now apply the
so-called effective field hypothesis which is the
main approximate hypothesis of many micromechanical methods:

 \noindent {\bf Hypothesis 1a, H1a)}. {\it Each inclusion $v_i$  has an ellipsoidal
form and is located
in the field (2.7$_2$)}
\BB
\overline {\bfsigma}_i({\bf y})
\equiv \overline {\bfsigma}({\bf x}_i)\ (\bfy\in v_i)
%(2.9)
\EE
{\it which is homogeneous
over the inclusion $v_i$}.

In some methods (such as, e.g., the MEFM) this basic hypothesis {\bf H1a} is complimented by a satellite hypothesis:

  \noindent{\bf Hypothesis 1b, H1b)}
{\it The perturbation introduced by the inclusion $v_i$
at the point ${\bf y}\notin v_i$ is defined by the relation}
\BB
\int{\bf \Gamma(y-x)}V_i({\bf x})\bfeta({\bf x})d{\bf x}=
\bar v_i{\bf T}_i{\bf (y-x}_i)
\bfeta_i.
%\eqno (2.10)$$
\EE
Hereafter $\bfeta_i\equiv \langle \bfeta({\bf x})V_i({\bf x})
\rangle _{(i)}$ is an average over
the  volume of the inclusion $v_i$ (but not over the ensemble),
$\langle(.)\rangle_i\equiv \langle\langle(.)
\rangle_{(i)}\rangle$, and ($\bfx\in v_i,\ \bfy\in v_j$)
\BB\bfT_i{\bf ( x \!\! - \!\! x}_i)\!\!=\!\!
\cases {-(\overline v_i)^{-1}\bfQ_i\
&{\rm for} ${\bf x} \in v_i,$\cr
(\overline v_i)^{-1}
\smallint {\bfGa(\bfx-\bfy)}V_i({\bf y})d{\bf y}
&{\rm for} $\bfx \not \in v_i,$\cr}\!\!,  \ \ \
{\bf T}_{ij}(\bfx_j-\bfx_i)=\langle{\bfT}_i(\bfy-\bfx_i)\rangle_{(j)}, %(2.11)
\EE
where the  tensor ${\bf Q}_i$ is associated with the
well-known Eshelby tensor by $\bfS_i=\bfI-\bfM^{(0)}{\bfQ_i}$.
For a homogeneous ellipsoidal inclusion $v_i$ the standard assumption (2.9)
(see, e.g., [6], [7]) % Buryachenko, 2007; Kanaun and Levin, 2008 )
yields  the assumption (2.10), otherwise the formula
(2.10) defines an additional assumption.
The tensors ${\bf T}_{ij}({\bf x}_i-{\bf x}_j)$,
proposed  by  Willis and Acton [25] %(1976)
for identical spherical inclusions have an analytical
representation for spherical inclusions of different size in an  isotropic
matrix (see for references [6]) %Buryachenko, 2007)
regardless of whether the inclusions are coated or uncoated.

According to hypothesis ${\bf H1a}$ and
 in view of the linearity of the problem  there exist
constant fourth and second-rank tensors ${\bf B}_i({\bf x}),\
{\bf R}_i{\bf (x)}$  and
${\bf C}_i{\bf (x)},\ {\bf F}_i{\bf (x)}$, such that
\BB\bfsigma ({\bf x})={\bf B}_i{\bf (x)}\overline
{\bfsigma} ({\bf x}_i)+{\bf C}_i{\bf (x)},\quad
\overline v_i \bfeta({\bf x})
={\bf R}_i{\bf (x)}\overline {\bfsigma} ({\bf x}_i)+{\bf F}_i{\bf (x)},
\quad {\bf x}\in v_i,
%\eqno (2.12)$$
\EE
where $v_i\subset v^{(i)}$ and
${\bf R}_i{\bf (x)}=\bar v_i{\bf M}_1^{(i)}({\bf x}){\bf B}_i{\bf
(x)},\
{\bf F}_i{\bf (x)}=\bar v_i[{\bf M}_1^{(i)}({\bf x)C}_i{\bf (x)}+
\bfbeta_1({\bf x})].$
According to {Eshelby}'s [23]
%(1957)
theorem there are the following relations between the
averaged tensors (2.12)
${\bf R}_i=\overline v_i
{\bf Q}_i ^{-1}({\bf I-B}_i)$,
${\bf F}_i
=-\overline v_i{\bf Q}_i^{-1}{\bf C}_i,$
where ${\bf g}_i\equiv \langle {\bf g(x)}\rangle _{(i)} \quad ({\bf g}$ stands for ${\bf
B,C,R,F})$.
It should be mentioned that the field $\overline{\bfsigma}(\bfx_i)$ can
vary with the location of the center $\bfx_i$ of the inclusion considered,
but the field $\overline{\bfsi}(\bfy)$ ($\bfy\in v_i)$ is homogeneous over the
inclusion $v_i$. Because of this the application of Eshelby's theorem
is correct.

For
example, for the homogeneous ellipsoidal domain $v_i$ with
\BB
{\bf M}_1^{(i)}({\bf x})=
{\bf M}_1^{(i)}={\rm const},\ \bfbeta_1^{(i)}({\bf x})=\bfbeta^{(i)}_1=
{\rm const}\quad {\rm at}
\ {\bf x}\in v_i, %(2.13)
\EE
 we obtain
\BB
{\bf B}_i=\left({{\bf I+Q}_i{\bf M}_1^{(i)}}\right)^{-1},\quad {\bf C}_i=
-{\bf B}_i{\bf Q}_i\bfbeta^{(i)}_1. %(2.14)
\EE
In the general case of coated inclusions $v_i$, the tensors
$\bfB_i(\bfx)$ and
$\bfC_i(\bfx)$ can be found by the transformation method by
 Dvorak  and {Benveniste} [26] %(1992)
(see for references and details [6], [27]).
% Buryachenko, 2007; Nogales and  B\"ohm, 2008).
% where the references of solution of the problem for coated ellipsoidal
% inclusion can be found].

              Using  hypothesis {\bf H1} (combining the hypotheses {\bf H1a} and {\bf H1b}), the system (2.7$_2$) for $k$ fixed inclusions
 with fixed values $\widetilde {\bfsigma}_{1,\ldots,k}(\bfx)
\ ({\bf x}\in v_i, \ i=1.\ldots,k)$  on the right-hand side
of the equations becomes algebraic when the  solution (2.12)
for one inclusion in the field $\overline {\bfsigma} ({\bf x}_i)
\quad (i=1,\ldots,k)$ is applied
\BB{\bf R}_i\overline {\bfsigma}({\bf x}_i) +{\bf F}_i=\sum^k_{j=1}{\bf Z}_{ij}
\Big\{{\bf R}_j\widetilde {\bfsigma}
 _{1,\ldots,k}(\bfx_j)+{\bf F}_j\Big\},
%(2.15)
\EE
where the  matrix ${\bf Z}^{-1}$   has the elements $({\bf Z}^
{-1})_{ij}\quad $
\BB({\bf Z}^{-1})_{ij}={\bf I}\delta_{ij}-(1-\delta_{ij})
{\bf R}_j{\bf T}_{ij}({\bf x}_i-
{\bf x}_j),\quad  (i,j=1,\ldots,n).
%(2.16)
\EE

\medskip
\noindent{\it 2.3 Closing effective field hypothesis and effective properties}
\medskip

Different methods can be employed (see for details [6]) to truncate the hierarchy (2.8) considered as a system of coupled equations. One begins with the last hierarchy item which has the most heterogeneities held fixed, because this equation does not depend on the other. The solution obtained presents the forcing term in the next equation up the hierarchy. The unconditionally average field is finally obtained by going step by step up the hierarchy.
For termination of the  hierarchy of statistical moment equations (2.8)
we will use the closing effective field hypothesis:

\noindent {\bf Hypothesis 2a, H2a)} {\it For a  sufficiently large $n$, the system (2.8) is closed
by the assumption}\\
%\noindent
$\langle \widetilde{\bfsigma}
_{1,\ldots,j,\ldots,n+1}(\bfx)\rangle _i
$ $=\langle \widetilde {\bfsigma}
_{1,\ldots,n}(\bfx)\rangle _i$,
{\it where the  right--hand--side of the equality  does
not contain  the index} $j\neq i\ (i=1,\ldots,n;\ j=1,\ldots,n+1;\
 {\bf x}\in v_i)$.

The hypothesis {\bf H2a}
rewritten  in terms of stresses $\bfsi(\bfx), \ (\bfx\in v_i)$ is a standard
closing assumption (see e.g. [1], [28])
%Khoroshun, 1978, 1987; Willis, 1981)
degenerating to the ``quasi-crystalline" approximation [16] %by Lax (1952)
at $n=1$ (see for analysis also Subsection 2.4).

In the framework of the hypothesis {\bf H1}, substitution  of the solution
(2.12),  and (2.15) (at $k=2$)  for binary interacting inclusions into the first equation of
 the system (2.8) at $n=1$ and  at the effective field hypothesis
{\bf H2a)} with the first order approximation:
\BB
\langle \widetilde {\bfsigma}_{i,q}(\bfx)\rangle _j=
\langle \overline {\bfsigma} ({\bf x})\rangle _j={\rm const}.\quad \bfx\in v_j (j=i,q).
%\eqno (2.17)$$
\EE
leads to the solution ($\bfx\in v_i$)
\BBEQ  \langle \overline {\bfsigma}  \rangle _i(\bfx)&=&
 \int {\bf T}_{q}({\bf x}-{\bf x}_q){\bf Z}_{qi}
\varphi (v_q,{\bf x}_q\vert ;v_i,{\bf x}_i)d{\bf x}_q
({\bf R}_i\langle \overline {\bfsigma}  \rangle _i+{\bf F}_i)\nonumber\\
&+&\int\left[{
 {\bf T}_{q}({\bf x}-{\bf x}_q){\bf Z}_{qq}
\varphi (v_q,{\bf x}_q\vert ;v_i;{\bf x}_i)-{\bfGa}({\bf x}-{\bf x}_q)
n^{(q)}}\right]
%\nonumber\\&\cdot&
({\bf R}_q \langle \overline {\bfsigma}  \rangle _q+{\bf F}_q)d{\bf x}_q,
%\eqno (2.18)
\EEEQ
where the
matrix elements ${\bf Z}_{qi},\ {\bf Z}_{qq}$ are nondiagonal elements
and diagonal ones of the binary interaction matrix ${\bf Z}$ (2.16) for the
two inclusions $v_q$ and $v_i$;
hereafter the conditional probability density $\varphi (v_q,{\bf x}_q\vert ;v_i;{\bf x}_i)$
and probability density $\varphi (v_q,{\bf x}_q)\equiv n^{(q)}$ are described in (I).
Averaging the result obtained (2.18) over the inclusion $v_i$
%at $N=1$ (a general case of $N$ is considered in Buryachenko, 2007)
yields the final representation for both the statistical average stress field and effective properties
\BBEQ
\langle  {\bfsigma}  \rangle _i(\bfx)
\!\!\!\!&=&\!\!\!\!
\bfB_i(\bfx)\bfR_i^{-1}[\sum_{j=1}^N{\bf Y}_{ij}({\bf R}_j \langle \bfsigma \rangle +\bfF_j)-\bfF_i]+\bfC_i(\bfx),\\ %(2.19)
\bfM^*\!\!\!\!&=&\!\!\!\!\bfM^{(0)}+\sum_{i,j=1}^N\bfY_{ij}\bfR_j n^{(i)},\ \ \
\bfbe^*= \bfbe^{(0)}+\sum_{i,j=1}^N\bfY_{ij}\bfF_j n^{(i)}  %(2.20)
\EEEQ
    where the matrix ${\bf Y}$ determines the action of the
surrounding inclusions on the considered one and has an inverse  matrix
${\bf Y}^{-1}$ given by
\BBEQ
({\bf Y}^{-1})_{ij}\!\!\!\!\! &=\!\!\!\!\!&\delta_{ij}\left[{{\bf I-R}_i
  \sum_{q=1}^N\int {\bf T}_{iq}({\bf x}_i-{\bf x}_q){\bf Z}_{qi}
  \varphi (v_q,{\bf x}_q\vert ;v_i,{\bf x}_i)d{\bf x}_q}\right]\nonumber\\
 \!\!\!\!\! &-\!\!\!\!\!& {\bf R}_i\int\left[{
  {\bf T}_{iq}({\bf x}_i-{\bf x}_q){\bf Z}_{qq}
  \varphi (v_q,{\bf x}_q\vert ;v_i,{\bf x}_i)-{\bf T}_i({\bf x}_i-
  {\bf x}_q)n^{(q)}}\right]d{\bf x}_q.
%\cr}\eqno (2.21)$$
\EEEQ
Buryachenko [6]
%(2007)
demonstrated that the MEFM includes in particular cases the well-known methods of mechanics of strongly heterogeneous media (such as the effective medium and the mean field methods).

\medskip
\noindent{\it 2.4 Quasi-crystalline approximation}
\medskip

Hypothesis {\bf H2a} rewritten in terms of stresses $\bfsi(\bfx)$ degenerates to the
``quasi-crystalline" approximation by Lax [16] %(1951, 1952)
which in our notations has two equivalent forms

\noindent {\bf Hypothesis 2b, H2b, ``quasi-crystalline" approximation}.
{\it It is supposed that the mean value of the effective field at a point
$\bfx\in v_i$ does not depend on the stress field  inside surrounding heterogeneities
$v_j\not = v_i$}:
\BB
\langle \overline{\bfsigma}_i({\bf x})\vert v_i,{\bf x}_i;v_j,{\bf x}_j \rangle = \langle \overline
{\bfsigma}_i \rangle ,\quad {\bf x}\in v_i, \ \ \ {\rm or}\ \ \
{\bf Z}_{ij}={\bf I}\delta_{ij}. %(2.22)
\EE
Therefore, the matrix ${\bf Y}^{-1}$ can be reduced to
(see  Ref. [29]) %Buryachenko and  Parton, 1990a)
\BB
({\bf Y}^{-1})_{ij}={\bf I}\delta_{ij}-{\bf R}_i\int[{\bf T}_{ij}({\bf x}_i-
{\bf x}_j)\varphi(v_j,{\bf x}_j\vert ; v_i,{\bf x}_i)-{\bf T}_i({\bf x}_i-
{\bf x}_j)n^{(j)}]d{\bf x}_j.
%\eqno (2.23)$$
\EE
% && The essence of hypothesis 3 is in detail described by Willis (1984).
The principal difference between
the hypotheses  (2.17) and (2.22) are discussed in Chapter 9 in [6]).

\noindent {\bf Note.}
It should be mentioned that the hypotheses {\bf H2a} and {\bf H2b} are not conceptually dependent on the hypothesis {\bf H1} and can be applied in general case even if the hypothesis {\bf H1} is violated (see for details Subsection 5.3).

\medskip
\noindent{\it 2.5 Hypothesis of ``{\it ellipsoidal symmetry}" of composite structure}
\medskip

To make further progress, the hypothesis of ``{\it ellipsoidal symmetry}" for the distribution of inclusions attributed to Willis [22] %(1977)
(see also Khoroshun [28], [31], %1974, 1978;
Buryachenko and Parton [29], %, 1990a;
Ponte Castaneda and Willis [32])
%, 1995)
is widely used:

\noindent {\bf Hypothesis 3, H3, ``ellipsoidal symmetry"}.
{\it The conditional probability density function $\varphi (v_{j},{\bf x}_j \mid ;v_{i},{\bf x}_{i})$ depends on $\bfx_j-\bfx_i$ only through the combination $\rho=|({\bf a}^0_{ij})^{-1} ({\bf x}_{j}-{\bf x}_{i})|$}:
\BB
\varphi (v_{j},{\bf x}_j \mid ;v_{i},{\bf x}_{i})
=h
(\rho ),\ \ \  \rho \equiv \mid
({\bf a}^0_{ij})^{-1} ({\bf x}_{j}-{\bf x}_{i})\mid
 %{2.24},
\EE
{\it where the matrix $({\bf a}^0_{ij})^{-1}$ (which is symmetric in the indexes $i$ and
$j$, ${\bf a}^0_{ij}={\bf a}^0_{ji}$)
defines the ellipsoid excluded volume $v^0_{ij}=\{\bfx:\ |({\bf a}^0_{ij})^{-1}\bfx|^2< 1\}$.}

A pair distribution function has ``ellipsoidal symmetry" but with an  ellipsoid shape
differing from the one that defines the inclusion shape. Although the assumed statistics may not be exactly realized in any particular composite, the results of effective moduli estimations are explicit and simple to use.
It is crucial for the analyst to be aware of their reasonable choice of the shape of ``ellipsoidal" spatial correlation of inclusion
location (see Chapter 18 in Ref. [6]). %Buryachenko, 2007).
For  spherical inclusions the
relation  (2.24) is realized for a statistical isotropy of the composite
structure. It is reasonable to assume that
$({\bf a}^0_{ij}) ^{-1}$ identifies
a matrix of affine transformation that transfers  the ellipsoid
$v_{ij}^0$ being the ``excluded volume" (``correlation hole") into a  unit sphere and, therefore, the representation of the matrix $\bfY_{ij}$ can be simplified:
\BB
({\bf Y}^{-1})_{ij}={\bf I}\delta_{ij}-{\bf R}_i\bfQ_{ij}^0,
%2.25
\EE
where $\bfQ_{ij}^0\equiv\bfQ(v_{ij}^0)$ is a constant for the ellipsoidal domain $v_{ij}^0$
 with the indicator function $V_{ij}^0$.
For the sake of simplicity of the subsequent calculation we will usually assume that the shape of ``correlation hole" $v^0_{ij}$ does not depend
on the inclusion $v_j$: $v^0_{ij}=v_i^0$ and $\bfQ_{ij}^0=\bfQ_i^0\equiv \bfQ(v_i^0)$.

Substitution of the representation (2.25) into Eqs. (2.19) and (2.20) completes the problem of effective properties estimations.
The hypothesis {\bf H3} is widely used for micromechanical structures described also by the indicator  function $V(\bfx|;v_1,\bfx_1;\ldots;v_n,x_n)$ (see, e.g. Willis [1], [22];
%1977, 1981, 1982;
Ponte Casta\~neda and  Willis [32]) %, 1995)
rather than only by the conditional probability density
$\varphi (v_{q},{\bf x}_q \mid ;v_{i},{\bf x}_{i})$ (2.24).

\noindent {\bf Note.} A popular point of view is that the hypothesis of ``{\it ellipsoidal symmetry}" (2.24) is exploited just for some simplification of the representation (2.23) reduced to (2.25). However, we will demonstrate in Section 5 that the destination of the hypothesis {\bf H3} is more fundamental and directed towards providing of conditions for applying of the hypothesis {\bf H1}. The use of the satellite hypothesis {\bf H3} has no sense without the hypothesis {\bf H1}.

\noindent {\bf Proposition 2.} If the hypotheses {\bf H1}, {\bf H2b}, {\bf H3} hold for the statistically homogeneous medium and homogeneous boundary conditions
then the effective properties $\bfM^*$ and $\bfbe^*$ do not depend on the size of the correlation hole $v_i^0$ and the conditional probability density (2.24).

We can reach this conclusion by simple analyses of the final representations (2.19), (2.20),
(2.23), and (2.25) as well as of analogous representations obtained by other methods in the framework of the hypotheses {\bf H1}, {\bf H2b}, and {\bf H3} (see, e.g., [1], [22], [28]).
%  Willis 1977, 1981;
% &&, 1982, 1984;
% Khoroshun 1978).
Here we keep in mind, first of all, the MEF which is equivalent for aligned identical ellipsoidal inclusions to the Mori-Tanaka [10] %(1973)
method.
Moreover, Markov [33] %(2001)
demonstrated that for homogeneous ellipsoidal inclusions, the estimations by the MEF coincide with the variational estimates obtained by
Ponte Castan\~eda and Willis [32] %(1995)
who also exploited the hypotheses {\bf H1}, {\bf H2b}, and {\bf H3}.
  It will be demonstrated in Section 6 that satisfiability of the hypotheses {\bf H2b} and {\bf H3} without {\bf H1} can lead to dependence of $\bfM^*$ and $\bfbe^*$ on both the size of the correlation hole $v_i^0$ and the binary correlation function.

\medskip
\noindent{\bf 3. A single inclusion subjected to inhomogeneous prescribed effective field}
\medskip

\noindent{\it 3.1 Operator representation for solution obtained by the VIE method}
\medskip

\setcounter{equation}{0}
\renewcommand{\theequation}{3.\arabic{equation}}

In the current section we will present a slightly modified solution of  a satellite problem (see [6] where additional references can be found) which is adapted for estimation of effective properties of composites in Section 6. Namely, let the inclusions  $v_i$ be fixed
and loaded by the inhomogeneous effective field
$\overline{\bfsi}_i(\bfx)$:
\BB
\bfsi(\bfx)=\overline{\bfsi}_i(\bfx)+\int{\bf \Gamma}(\bfx-\bfy)V_i(\bfy)
\bfeta(\bfy)d\bfy, %(3.1
\EE
A known quadrature method for obtaining an approximate solution
of Eq. (3.1) is to  evaluate the volume integrals with a Gauss
quadrature formula. Then the corresponding equations at the Gauss points will contain a singular
term that results when the field point $\bfx$ and source point $\bfy$ coincide: $\bfx=\bfy$
(3.1). The difficulties with the   troublesome singularities can be avoided if a rearrangement of Eq. (3.1) is performed
in the spirit of a subtraction technique used in the modified quadrature method (see, e.g.,
[34]) %Delves and Mohamed, 1985)
\BBEQ
\bfeta ({\bf x})&=&{\bf M}_1(\bfx)\overline{\bfsi}(\bfx)+\bfbe_1(\bfx)+
{\bf M}_1(\bfx) \int V_{i0}(\bfy){\bf \Gamma (x-y)}d{\bf y}
\bfeta(\bfx)\nonumber\\
&+&{\bf M}_1(\bfx)\int {\bf \Gamma (x-y)}
\big[\bfeta({\bf y})-V_{i0}(\bfy)\bfeta(\bfx)\big]
 d{\bf y,} \quad \bfx\in v_i,
%\eqno (3.2)
\EEEQ
where $v_i$, perhaps, is not an ellipsoid.
The  equation is valid for any domain $\bfx\in v_{i0}$
with the indicator function $V_{i0}(\bfx)$.
We assume that the first integral in (3.2) is easily computable
\BB-\bfQ_{i0}(\bfx)=\int V_{i0}(\bfy){\bf \Gamma (x-y)}d{\bf y}. %(3.3)
\EE
In a general case of the inclusion shape $v_i$ the function $\bfQ_{i0}(\bfx)$
can be found numerically, e.g. by finite element analysis
(see, e.g.,  [35]).   %Buryachenko and Tandon, 2004).
For ellipsoidal domain $v_{i0}\supset v_i$,
the first integral on the
right-hand-side of (3.2) is known and is associated with the
well-known Eshelby tensor by $(\bfx\in v_i\subset v_{i0},\ \bfy\in v_{i0})$
\BB
{\bf S}_{i0}={\bf I- M}^{(0)}{\bf Q}_{i0},
\quad {\bf Q}_{i0}\equiv -\bar v_{i0} \langle
{\bf \Gamma(x-y)} \rangle^0 _i=
{\rm const.}%(3.4)
\EE
Hereafter
\BB
{\bf g}_i\equiv  \langle {\bf g(y)} \rangle _{i}=
\bar v_i^{-1}\int {\bf g}(\bfy)V_i(\bfy)d\bfy,\quad
{\bf g}_{i0}\equiv  \langle {\bf g(y)} \rangle^0 _{i}=
(\bar v_{i0})^{-1}\int {\bf g}(\bfy)V_{i0}(\bfy)d\bfy
%(3.5)
\EE
denotes averaging of some tensor ${\bf g}(\bfy)$ over
the volume of the regions $\bfy\in v_i\subset v_{i0}$ and $\bfy\in v_{i0}$,
respectively;
in so doing, $\bfx\in v_i$ is fixed.
The assumption of an  ellipsoidal shape of the domain $v_{i0}$ was used only
to obtain  analytical representation of the integral (3.4).
This is because  the tensor $\langle{\bf \Gamma (x-y)}\rangle_{i0}$ is  homogeneous
for ${\bf x}\in v_i,\ \bfy\in v_{i0}$  for an ellipsoid.
For nonellipsoidal inclusions $v^{\rm n-e}$ one could assume that in some
parts of the region $v_i^{\rm c}\subset v_i$ the properties
$ {\bf M}_1({\bf x})\equiv {\bf 0}$, $\beta_1({\bf x})\equiv {\bf 0}$, i.e.
it is sufficient to replace a real nonellipsoidal inclusion
$v^{\rm n-e}=v_i\setminus v_i^{\rm c}$
by a fictitious ellipsoid (with smallest possible  volume) and call it the
inclusion $v_i$ with a ``coating" $v^{\rm c}_i$.
In so doing, at the estimation of the second integral in (3.2)
we keep in mind that $\smallint \bfGa (\bfx-\bfy)
(V_i(\bfy)-V_{i0}(\bfy))d\bfy=\bfQ_{i0}-\bfQ_i(\bfx) \ (\bfx\in v_i)$.
With  the nonessential restriction on the shape of
the inclusion $v_i$ mentioned
above, we can consider without loss of generality
an ellipsoidal inclusion $v_{i0}=v_i$;
in so doing $ {\bf M}_1({\bf y})\equiv {\bf 0}$, $\beta_1({\bf y})\equiv {\bf 0}$
at $\bfy\in v_i^{\rm c}\subset v_i$.
Then Eq. (3.2) can be rewritten in the equivalent compact form
\BB
\bfeta ({\bf x})=\overline{\bfeta}_i(\bfx)+
\int \bfK_i (\bfx,\bfy)
\big[\bfeta({\bf y})-\bfeta(\bfx)\big]
 d{\bf y,} \quad \bfx\in v_i,
%\eqno (3.6)
\EE
where $\overline{\bfeta}_i(\bfx)={\bf E}_i(\bfx)\overline{\bfsi}(\bfx)+
{\bf H}_i(\bfx)$, $(\bfx\in v_i$) is called
the effective strain polarization tensor in the inclusion $v_i$,
and (no sum on $i$)
\BBEQ
{\bf K}_i(\bfx,\bfy)&=&{\bf E}_i(\bfx)
{\bf \Gamma}(\bfx-\bfy)V_i(\bfy), \\%(3.7)
{\bf E}_i(\bfx)&=&{\bf M}_1(\bfx)
[{\bf I}+{\bf Q}_{i0}(\bfx){\bf M}_1(\bfx)]^{-1},\\%(3.8)
{\bf H}_i(\bfx)&=&[{\bf I+M}_1(\bfx){\bf Q}_{i0}(\bfx)]^{-1}\bfbe_1(\bfx).
%(3.9)
\EEEQ
We rewrite Eq. (3.6)
in symbolic form:
\BB
\bfeta=\overline{\bfeta}_i+\bfcK_i\bfeta,
%(3.10)
\EE
where \BB
\big(\bfcK_i\bfet\big)(\bfx)=\int\bfcK_i(\bfx,\bfy)\bfet(\bfy)d\bfy
%(3.11)
\EE
defines the integral operator $\bfcK_i$ with the kernel formally represented as
\BB
\bfcK_i(\bfx,\bfy)={\bf K}_i(\bfx,\bfy)-\delta(\bfx-\bfy)\int
V_i(\bfz){\bf K}_i(\bfx,\bfz)d\bfz .%(3.12)
\EE
We formally write the solution of Eq. (3.10)
as
\BB
\bfet=\bfcL_i\!*\!\overline{\bfet}_i,
%(3.13)
\EE
where the inverse operator $\bfcL_i=({\bf I}-\bfcK_i)^{-1}$ will be
constructed by
the iteration method based on the  recursion formula
\BB
\bfeta^{[k+1]}=\overline{\bfeta}_i+\bfcK_i\bfeta^{[k]}
%(3.14)
\EE
to construct a sequence of functions $\big\{\bfeta^{[k]}\big\}$ that can be
treated as an approximation of the solution of Eq. (3.10).
We presented the point Jacobi (called also Richardson and point total-step)
iterative scheme for ease of calculations. The details of the real iteration method used
for the solution of Eq. (3.14) will be presented in Section 6.
Usually the driving term of this equation is used as an initial approximation:
\BB
{\bfet}^{[0]}
(\bfx)=\overline{\bfet}_i(\bfx),
%(3.15)
\EE
which is exact for a homogeneous ellipsoidal inclusion subjected to
 remote homogeneous stress field $\overline{\bfsi}(\bfx)\equiv \overline{\bfsi}={\rm const.}$
The sequence $\{{\bfet}^{[k]}\}$ (3.14) with arbitrary continuous
$\bfet^{[0]}(\bfx)$ converges to a unique solution
 ${\bfet}$  if the norm of the integral operator  $\bfcK_i$
turns out to be  small ``enough" (less than 1), and the problem is reduced to the computation of
 the integrals involved, the density of which is given. In effect the
iteration method (3.14) transforms the integral equation problem
(3.14) into the linear algebra problem in any case.

We will introduce the linear operators $\bfcL^{\sigma}$ and $\bfcL^{\eta}$ describing a perturbation of the stress field inside and outside the inclusion $v_i$ ($\bfx\in R^d$)
\BBEQ
\!\!\!\!\!\!\!\!\int{\bf \Gamma}(\bfx-\bfy)V_i(\bfy)
\bfeta(\bfy)d\bfy\!\!\!\!&=&\!\!\!\!\bfsi(\bfx)-\overline{\bfsi}_i(\bfx)\equiv
\bfcL_i^{\sigma}(\overline{\bfsi}_i)(\bfx)\equiv \bfcL^{\eta}_i(\bfet)(\bfx),\\ %(3.16)
\bfcL_i^{\sigma}(\overline{\bfsi}_i)(\bfx)
\!\!\!\!&=&\!\!\!\! \int{\bf \Gamma}(\bfx-\bfy)
\bfcL_i*(\bfE_i\overline{\bfsi}+\bfH_i)(\bfy)V_i(\bfy) d\bfy, \\%(3.17)
\bfcL_i^{\eta}(\bfet)(\bfx)
\!\!\!\!&=&\!\!\!\! \int{\bf \Gamma}(\bfx-\bfy)\bfet(\bfy)V_i(\bfy)
d\bfy.  %(3.18)
\EEEQ
The right-hand side of Eqs. (3.17) and (3.18) can be also estimated in the spirit of subtraction technique (3.2) according to the next scheme
\BBEQ
\bfcL(\bfg)(\bfx)&\equiv&\int\bfGa(\bfx-\bfy)\bfg(\bfy)V_i(\bfy)d\bfy\nonumber\\
\!\!\!\!&=&\!\!\!\!\bfQ_i(\bfx)\bfg(\bfx)+\int\bfGa(\bfx-\bfy)[\bfg(\bfy)-\bfg(\bfx)]V_i(\bfy)d\bfy,\\ %(3.19)
\bfcL(\bfg)(\bfx)\!\!\!\!&=&\!\!\!\!\bfQ_i(\bfx_m)\bfg(\bfx_m)
+\int\bfGa(\bfx-\bfy)[\bfg(\bfy)-\bfg(\bfx_m)]V_i(\bfy)d\bfy, %(3.20)
\EEEQ
for $\bfx\in v_i$ and $\bfx\not\in v_i$, respectively; here $\bfx_m=\arg\min_y|\bfx-\bfy|$ ($\bfy\in v_i,\
\bfx\not\in v_i$),
$\bfcL=\bfcL_i^{\eta},\bfcL_i^{\sigma}$, $\ \bfg=\overline{\bfsi}_i, \bfet$,
 and the tensor $\bfQ_i(\bfx)$ is defined analogously to Eq. (3.3).

We constructed the solution (3.16) for a perturbation of the stress field inside and outside the inclusion $v_i$ in the operator form obtained by the method of volume integral equation (VIE) for an arbitrary effective field $\overline{\bfsi}_i(\bfx),\ (\bfx\in v_i)$.
 However, this operator could be created by any another numerical method such as, e.g. the finite element analysis (FEA). The main difficulty in such a case is a generation of prescribed effective field $\overline{\bfsi}_i(\bfx),\ (\bfx\in v_i)$ which will be considered in the next subsection.

\medskip
\noindent{\it 3.2  Creation of prescribed stresses by the FEA}
\medskip

Construction of the operator $\bfcL^{\sigma}$ (3.16) anticipates a creation of prescribed effective field $\overline{\bfsi}_i(\bfx),\ (\bfx\in v_i)$ in the absence of the inclusion $v_i$. In the case of the FEA employment, it can be done by prescribing of either some boundary condition at the boundary of a large sample or some eigenstress inside this sample. We will only consider the second way.

The problem is to find a fictitious $\bfbe_1(\bfx)$ generating a prescribed stress
$\overline{\bfsi}(\bfx)$ in an arbitrary fictitious ellipsoidal inclusion $\bfx\in v_0$
(which has no connection with the correlation hole $v_{ij}^0$) with the elastic modulus $\bfM(\bfx)\equiv \bfM^{(0)}$ and an indicator function $V_0$.
\BB
\overline{\bfsi}(\bfx)=\int \bfGa(\bfx-\bfy)\bfbe_1(\bfy)V_0(\bfy)d\bfy.  \label{3.21}
\EE
The equation (\ref{3.21}) can be recast in the form
\BB
\bfbe_1(\bfx)=-\bfQ_0^{-1}\overline{\bfsi}(\bfx)+\bfQ_0^{-1}\int\bfGa(\bfx-\bfy)[\bfbe_1(\bfy)-\bfbe_1(\bfx)]V_0(\bfy)d\bfy  \label{3.22}
\EE
Except for notations,
the Fredholm integral equation of the second kind ({\ref{3.22})
coincides with the direct equation for estimation of stresses $\bfsi(\bfx)$ produced by the field $\overline{\bfsi}(\bfx)$ inside the heterogeneity ($\bfx\in v_0$)
 \BB
\bfsi(\bfx)=\bfB_0\overline{\bfsi}(\bfx)+\bfB_0\int\bfGa(\bfx-\bfy)[\bfM_1(\bfy)\bfsi(\bfy)-\bfM_1(\bfx)\bfsi(\bfx)]V_0(\bfy)d\bfy  \label{3.23}
\EE
Indeed, Eq. (\ref{3.22}) is reduced to Eq. (\ref{3.23}) (with $\bfbe_1\equiv {\bf 0}$) by the replacement of the notations: $\bfbe_1\to\bfM_1\bfsi$, $-\bfQ_0^{-1}\overline{\bfsi}\to
\bfM_1\bfB_0\overline{\bfsi}$, $\bfQ_0^{-1}\bfGa\to \bfM_1\bfB_0\bfGa$, where $\bfB_0=\bfB_0(v_0)$ is defined analogously to (2.14).

In the case of the successive approximations method, we need to evaluate the right-hand side of the equation
\BB
\bfbe_1^{[n+1]}(\bfx)=-\bfQ_0^{-1}\overline{\bfsi}(\bfx)+\bfQ_0^{-1}\int\bfGa(\bfx-\bfy)[\bfbe_1^{[n]}(\bfy)-\bfbe_1^{[n]}(\bfx)]V_0(\bfy)d\bfy  \label{3.24}
\EE
with the usual use of the driving term as an initial approximation
\BB
\bfbe_1^{[0]}(\bfx)=-\bfQ_0^{-1}\overline{\bfsi}(\bfx). \label{3.25}
\EE
The volume integral equation (\ref{3.21}) is reduced to the regular representation, which
has no singularities and can be also presented in the form adopted for using of the FEA
\BB
\bfbe^{[n+1]}(\bfx)=\bfQ_0^{-1}\overline{\bfsi}(\bfx)+\bfbe_1^{[n]}(\bfx)
-\bfQ_0^{-1}\bfcR\!*\!\bfbe_1^{[n]}(\bfx)  \label{3.26}
\EE
where the operator
\BB
\bfcR\!*\!\bfbe_1^{[n]}(\bfx)= \int\bfGa(\bfx-\bfy)\bfbe_1^{[n]}(\bfy)
V_0(\bfy)d\bfy\label{3.27}
\EE
presents the stresses produced by the intermediate eigenstresses $\bfbe_1^{[n]}(\bfx)$ in the inclusion $\bfx\in v_0$ (see Eq. (\ref{3.21})). Obviously, the mentioned stresses can be easy estimated by the FEA.
Thus, an operator representation of the solution
(\ref{3.21})
\BB
\bfbe(\bfx)=\bfGa^{-1}\!*\!\overline{\bfsi}(\bfx)\equiv\int
\bfGa^{-1}(\bfx-\bfy)\overline{\bfsi}(\bfy)V_0(\bfy)d\bfy \label{3.28}
\EE
can be considered as found by the FEA (or by any other numerical method providing a solution of the regular integral Eq. (\ref{3.22}). The next step for FEA utilization is obvious. We introduce the real inclusion $v_i$ into the fictitious ellipsoid $v_0$ (such that all desirable area for the stress estimation is placed inside $v_0$) and estimate the real stresses $\bfsi(\bfx)$ which can be considered as found in Eq. (3.16).

The VIE and FEA methods have a series of advantages
and disadvantages (considered, e.g., in Ref. [6]),
%by Buryachenko, 2007),
and it is crucial for the analyst to be aware of their range of applications.

\bigskip
\noindent{\bf 4. Estimation of both the effective field and effective elastic moduli}
\medskip

\setcounter{equation}{0}
\renewcommand{\theequation}{4.\arabic{equation}}

%\noindent{\it 4.1 General integral equation}
%\medskip

The new general integral equation (2.1)
%\BBEQ
%\bfsi ({\bf x})=\langle \bfsi\rangle +\int
% [{\bf \Gamma} (\bfx-\bfy) \bfeta({\bf y})-\lle\bfGa(\bfx-\bfy)\bfeta(\bfy)
%\rle(\bfy)]d\bfy,
% \eqno (4.1)$$
%\EEEQ
can be rewritten in terms of the operator representation $\bfcL^{\sigma}$ (3.16)
\BB
\bfsi(\bfx)=\lle\bfsi\rle(\bfx)+\int[\bfcL^{\eta}(\bfet)(\bfx)-
\lle\bfcL^{\eta}(\bfet)\rle(\bfx)]d\bfy %(4.1),
\EE
while conditional averaging of Eqs. (2.7$_2$) and (2.7$_3$) leads to the following representation for the mean of the effective field in the fixed inhomogeneity $\bfx\in v_i$
\BBEQ
\lle\overline{\bfsi}\rle_{i}(\bfx)\!\!\!\!&=&\!\!\!\!\lle\bfsi\rle(\bfx)+\int[
\bfcL_q^{\sigma}(\lle\overline{\bfsi}|;v_i,\bfx_i\rle_{q})(\bfx)
\varphi(v_q,\bfx_q|;v_i,\bfx_i)
%\nonumber\\\!\!\!\!&-&\!\!\!\!
-\bfcL_q^{\sigma}(\lle\overline{\bfsi}\rle_{q})(\bfx)n^{(q)}(\bfx_q)]d\bfx_q,%(4.2)
\EEEQ
where $\lle\overline{\bfsi}|;v_i,\bfx_i\rle_{q}\equiv
\lle\overline{\bfsi}|;v_i,\bfx_i\rle_{q}(\bfy)$ is a conditional statistical average of $\overline{\bfsi}(\bfy)$ varying along the fixed heterogeneity $\bfy\in v_q$ at the fixed $v_i$  while $\lle\overline{\bfsi}\rle_{q}\equiv
\lle\overline{\bfsi}\rle_{q}(\bfy)$ is a statistical average $\overline{\bfsi}(\bfy)$ inside the heterogeneity $\bfy\in v_q$. No confusion will arise hereafter in definition of the operator $\bfcD$ ($\bfcL_q^{\sigma},\ \bfcL_q^{\eta})$ with the kernel $\bfcD(\bfx,\bfy)$ on the inhomogeneous functions $\bfg(\bfx)$
(e.g., $\bfg(\bfy)=\lle\overline{\bfsi}\rle_{q}(\bfy), \ \ \bfy\in V_k;
V_k=V_i,V_q,V_i+V_q$)
\BB
\bfcD(\bfg)(\bfx)=\int \bfcD(\bfx,\bfy)\bfg(\bfy)V_k(\bfy)d\bfy.
%(4.3)
\EE
The operator $\bfcD$ is reduced to the tensor $\bfD(\bfx)$ on the constant functions $\bfg(\bfx)=\bfg\equiv$const ($\bfx\in V_k$)
\BB
\bfcD(\bfg)(\bfx)=\bfD(\bfx)\bfg,\ \ \bfD(\bfx)=\int\bfcD(\bfx,\bfy)V_k(\bfy)d\bfy.
%(4.4)
\EE

The integral in the right-hand side of Eq. (4.2) can be decomposed as
\BBEQ
\!\!\!\!\!\!\!\!\lle\overline{\bfsi}\rle_{i}(\bfx)\!\!\!\!&=&\!\!\!\!\lle\bfsi\rle(\bfx)+J,\ \ \ J=J_1+J_2+J_3, \ \ \ {\rm where}\\ %(4.5)
J_1\!\!\!\!&=&\!\!\!\!\int[\bfcL_q^{\sigma}(\lle\overline{\bfsi}|;v_i,\bfx_i\rle_{q})(\bfx)
-\bfcL_q^{\sigma}(\lle\overline{\bfsi}\rle_{q})(\bfx)]
\varphi(v_q,\bfx_q|;v_i,\bfx_i)d\bfx_q, \\%(4.6)
J_2\!\!\!\!&=&\!\!\!\!\int\bfcL_q^{\sigma}(\lle\overline{\bfsi}\rle_{q})(\bfx)
[\varphi(v_q,\bfx_q|;v_i,\bfx_i)-n^{(q)}(\bfx_q)][1-V^0_{iq}(\bfx_q)]d\bfx_q, \\%(4.7)
J_3\!\!\!\!&=&\!\!\!\!-\int\bfcL_q^{\sigma}(\lle\overline{\bfsi}\rle_{q})(\bfx)n^{(q)}(\bfx_q)
V^0_{iq}(\bfx_q)d\bfx_q. %(4.8)
\EEEQ
The absolutely convergent integral in Eq. (4.2) is decomposed in Eq. (4.5) just for subsequent presentation obviousness; because of this, the absolute convergences
of integrals $J_1,J_2$ and $J_3$ are not considered.

In the framework of the quasi-crystalline approximation (2.22) $(\bfy\in v_q)$
\BBEQ
\lle{\bfsi}|;v_i,\bfx_i\rle_{q}(\bfy)\!\!&=&\!\!
\lle{\bfsi}\rle_{q}(\bfy), \\ %(4.9)\ \ \
\lle\overline{\bfsi}|;v_i,\bfx_i\rle_{q}(\bfy)\!\!&=&\!\!
\lle\overline{\bfsi}\rle_{q}(\bfy),  %(4.10)
\EEEQ
and Eq. (4.5) is simplified ($J=J_2+J_3$)
\BB
\lle\overline{\bfsi}\rle_{i}(\bfx)=\lle\bfsi\rle(\bfx)+\int
\bfcL^{\sigma}_q(\lle\overline{\bfsi}\rle_{q})(\bfx)
[\varphi(v_q,\bfx_q|;v_i,\bfx_i)-n^{(q)}(\bfx_q)]d\bfx_q, %(4.11)
\EE
For statistically homogeneous media when $n(\bfx_q)=n^{(q)}\equiv$const.,
it is logical to assume the acceptance of the additional hypothesis of ``ellipsoidal symmetry" (2.24), which leads to the same simplification as in Subsection 2.4 ($J_2={\bf 0}$)
\BB
\int\bfcL^{\sigma}_q(\lle\overline{\bfsi}\rle_{q})(\bfx)
[\varphi(v_q,\bfx_q|;v_i,\bfx_i)-n^{(q)}(\bfx_q)][1-V^0_{iq}(\bfx_q)]d\bfx_q={\bf 0}, %(4.12)
\EE
which can be presented in an equivalent form exploiting Green's function
\BB
\!\int\!\bfGa(\bfx\!-\!\bfy)[\lle\bfM_1\bfsi\rle_{q}(\bfy)+\bfbe_1(\bfy)]
[\varphi(v_q,\bfx_q|;v_i,\bfx_i)\!-\!n^{(q)}]V_q(\bfy)[1\!-\!V^0_{iq}(\bfy)]d\bfy\!=\!{\bf 0}. %(4.13)
\EE
Then Eq. (4.5) leads to ($J=J_3$)
\BB
\lle\overline{\bfsi}\rle_{i}(\bfx)=\lle\bfsi\rle(\bfx)-\int n^{(q)}
\bfcL^{\sigma}_q(\lle\overline{\bfsi}\rle_{q})(\bfx)V^0_i(\bfx_q)d\bfx_q. %(4.14)
\EE
The accuracies of the assumptions (4.9), (4.10) and (4.12) will be estimated in Section 6. However,
we will perform subsequent solution of Eq. (4.11) rather than Eq. (4.14) by keeping in mind that the hypothesis {\bf H3} (2.24) is accepted.

Obviously, the regular integral equation (4.11) has no singularities and can be solved by a direct quadrature method with formal representation of the solution
\BB
\lle\overline{\bfsi}\rle_{i}(\bfx)=\bfcT_i\!*\!\lle\bfsi\rle(\bfx) %(4.15)
\EE
Although the direct quadrature method usually causes no problems of accuracy,
for a large number of unknown variables $N$ its $O(N^3$) cost dependence can lead
to surprisingly long computing time. The obvious way of reducing this cost is
to construct an iterative scheme which will be considered now.
Namely, Eq. (4.11) will be solved by the iteration method when the initial constant effective stress $\lle\overline{\bfsi}\rle_{q}(\bfx)=\lle\overline{\bfsi}\rle_{q}\equiv$const. is estimated from the classical approach (2.18) and (2.24). Indeed, for $\lle\overline{\bfsi}\rle_{q}\equiv$const., the statistical average of stresses inside the inhomogeneity $v_q$ is found
to be inhomogeneous [see Eq. (2.12)]
\BB
\lle\bfsi\rle_{q}(\bfx)=\bfB_q(\bfx)\lle\overline{\bfsi}\rle_{q}+\bfC_q(\bfx),
\ \ \ \lle\bfet\rle_{q}(\bfx)=\bfR^v_q(\bfx)\lle\overline{\bfsi}\rle_{q}+\bfF^v_q(\bfx), %(4.16)
\EE
 in a general case of inhomogeneity of $v_q$; here $\bfR^v_q=\bar v_q^{-1}\bfR_q,\
\bfF^v_q=\bar v_q^{-1}\bfF_q$. In such a case, the right-hand side of Eq. (4.11) generates inhomogeneous field $\lle\overline{\bfsi}\rle_{i}(\bfx)$. For elimination of this difficulty, we will use the additional condition of the effective field hypothesis {\bf H1b} (2.10), when a perturbation introduced by the inhomogeneity $v_q$
is defined by the strain polarization tensor $\lle\bfet\rle_q$ averaged over the
volume $v_q$ ($\bfx\in R^d$)
 \BB
\bfcL^{\sigma}_q(\lle\overline{\bfsi}\rle_{q})(\bfx)=\bar v_q\bfT_q(\bfx-\bfx_q)\lle\bfet\rle_{q},
%(4.17)
\EE
Then Eq. (4.11) in the framework of the hypothesis {\bf H3} (2.24) is reduced to the classical representation for the effective field
$\lle\overline{\bfsi}\rle_{i}(\bfx)$
\BB
\lle\overline{\bfsi}\rle_{i}(\bfx)=\lle\bfsi\rle
+\bfQ_i^0(\bfx)\sum_{q=1}^n c^{(q)}\lle\bfet\rle_{q}, %(4.18)
\EE
which is homogeneous just for an additional assumption of an ellipsoidal
shape of the excluded volume $v_{i}^0$. Combining the averaged Eqs. (4.16) and (4.18) leads to the final representations for the averaged tensors of both the  effective field and
stain polarization
\BBEQ
\!\!\!\!\!\!\!\!\!\!\!\!\!\!\!\!\lle\overline{\bfsi}\rle_i^{[0]}\!\!\!\!&=&\!\!\!\!\lle\bfsi\rle\!+\!\bfQ^0_i[\bfI\!-\!\lle\bfR^v\bfQ^0V\rle]^{-1}
(\lle\bfR^vV\rle\lle\bfsi\rle\!+\!\lle\bfF^vV\rle), \\ %(4.19)
\!\!\!\!\!\!\!\!\!\!\!\!\!\!\!\!\bar v_i \lle{\bfsi}\rle_i^{[0]}(\bfx)\!\!\!\!&=&\!\!\!\!\bfB_i(\bfx)\lle\bfsi\rle\!+\!\bfC_i(\bfx)\!+\!\bfB_i(\bfx)\bfQ^0_i[\bfI\!-\!\lle\bfR^v\bfQ^0V\rle]^{-1}\!
(\lle\bfR^vV\rle\lle\bfsi\rle\!+\!\lle\bfF^vV\rle),\\ %(4.20)
\!\!\!\!\!\!\!\!\!\!\!\!\!\!\!\!\lle{\bfet}\rle_i^{[0]}(\bfx)\!\!\!\!&=&\!\!\!\!\bfR_i(\bfx)\lle\bfsi\rle\!+\!\bfF_i(\bfx)\!+\!\bfR_i(\bfx)\bfQ^0_i[\bfI\!-\!\lle\bfR^v\bfQ^0V\rle]^{-1}\!
(\lle\bfR^vV\rle\lle\bfsi\rle\!+\!\lle\bfF^vV\rle), %(4.21)
\EEEQ
 which will be considered as the initial approximation of the next equations ($\bfx\in v_i,\ \bfy\in v_q)$
\BBEQ
\!\!\!\!\lle\overline{\bfsi}\rle_{i}^{[n+1]}(\bfx)\!\!\!\!&=&\!\!\!\!\lle\bfsi\rle+\int
\bfcL^{\eta}_q(\lle{\bfet}\rle_{q}^{[n]})(\bfx)
[\varphi(v_q,\bfx_q|;v_i,\bfx_i)-n^{(q)}(\bfx_q)]d\bfx_q,\\ %(4.22)
\!\!\!\!\lle\bfet\rle_{q}^{[n+1]}(\bfy)\!\!\!\!&=&\!\!\!\!\bfcR_q\!*\!\lle\overline{\bfsi}\rle_{q}^{[n+1]}(\bfy)+\bfF_q^v(\bfy), %(4.23)
\EEEQ
where $\bfcR_q=\bfM_1(\bfI+\bfcL^{\sigma}_q)$ and Eq. (4.22) in the case of the assumption
(4.12) is reduced to the following one
\BB
\lle\overline{\bfsi}\rle_{i}^{[n+1]}(\bfx)=\lle\bfsi\rle-\int
\bfcL^{\eta}_q(\lle{\bfet}\rle_{q}^{[n]})(\bfx)
n^{(q)}(\bfx_q)V^0_i(\bfx_q)d\bfx_q.\\ %(4.24)
\EE

The system (4.22) and (4.23) can be formally presented in an operator form $\lle\bfet\rle
(\bfx)=\lle\bfsi\rle+\bfcK(\lle\bfet\rle)(\bfx)$ (the indexes are dropped for simplicity).
It suggests the Neumann series form for the  solution ${\bfet}$ of (4.22) and (4.23)
[compare with the solution (4.15)]
\BB
\lle{\bfet}\rle_i(\bfx)\equiv\lim_{n\to \infty}\lle\bfet^{[n]}\rle_i(\bfx)=\bfR_i^*(\bfx)
\lle\bfsi\rle+\bfF_i^*(\bfx), %(4.25)
\EE
which yields the final representations for the effective properties
\BB
\bfM^*=\bfM^{(0)}+\lle\bfR^*V\rle,\ \ \
\bfbe^*=\bfbe^{(0)}+\lle\bfF^*V\rle. %(4.26)
\EE
A convergence of the sequence $\lle\bfet^{[n]}\rle_i(\bfx)$ (4.25) is analyzed analogously to the sequence (3.14).
%Thus, the choice of the initial approximation in the form of the MEF is not

In Eq. (4.26) we used an obvious connection between the phase average $\lle \bfg V\rle$
($\bfg=\bfsi,\bfep,\bfet$)
and the averages inside the representative inclusions $v_k\in v^{(k)}$ ($k=1,\ldots,N$)
\BB
\lle \bfg V\rle=\sum_{k=1}^N c^{(k)} \lle\bfg\rle_k, %(4.27)
\EE
which is only fulfilled for statistically homogeneous media subjected to the homogeneous boundary conditions. If any of these conditions is broken then it is necessary to consider a generalization of Eq. (4.27) in the form of Eq. (3.29I). However, the mentioned class of nonlocal problems is beyond the scope of the present paper.

\bigskip
\noindent{\bf 5. Qualitative analysis of some basic hypotheses and propositions}

\medskip
\noindent{\it 5.1  Analysis of the proposition 1 and hypothesis {\bf H1}}
\medskip

\setcounter{equation}{0}
\renewcommand{\theequation}{5.\arabic{equation}}
%\textcolor{blue}{
The hypothesis {\bf H1} is widely used (explicitly or implicitly) for the majority of the methods of micromechanics even if the term ``effective field hypothesis" is not indicated. For example, Buryachenko [6]
%(2007)
demonstrated that hypothesis {\bf H1} is exploited in the effective medium method, generalized self-consistent method, differential methods, Mori-Tanaka method, the MEFM, conditional moments method, variational methods, and others. These are a lot of other methods using the hypothesis {\bf H1} differ one from one another by some additional specific assumptions.

It should be mentioned, that the domain of the operator $\bfcL^{\eta}_q(\lle{\bfet}\rle_{q}^{[n]})(\bfx)$ (3.18) is a whole space $\bfx\in R^d$, and, because of this,
some points of the area $\bfx\in v_i$ in Eq. (4.22) can be uncovered by the heterogeneities $v_q$ and, therefore, the effective stress
$\lle\overline{\bfsi}\rle_i^{[n+1]}(\bfx)$ (4.22) will depend on the stress perturbations
$\bfcL^{\sigma}_q(\lle\bfet\rle_q^{[n]})(\bfx)$ in the vicinity $\bfx\in v_i\setminus v_q$ of the area $v_q$ rather than only on stress distributions in the inhomogeneity $v_q$ and
${\max}_{x}|\bfx-\bfx_q|=3a$ ($\bfx\in v_i\setminus v_q$) for the identical spherical inhomogeneities of the radius $a$ with an isotropic statistically distribution of their centers. Thus, we obtain a fundamental conclusion that effective moduli in general  depend not only on the stress distribution inside the inhomogeneities but also on the stresses in the vicinities
of inhomogeneities (compare with the proposition 1). However, if our estimations utilize Eq. (2.2) containing only average strain polarization $\bfet_q$ [rather than $\lle\bfGa(\bfx-\bfy)\bfeta(\bfy)\rle_q$ in Eq. (4.1)] as a renormalizing item then an influence of stresses in the vicinities of inhomogeneities is degenerated. At the same time, using Eq. (4.1) leads to the necessity of evaluation of stresses in the inhomogeneity vicinities even for a statistically homogeneous field of ellipsoidal homogeneous inclusions (it will be quantitatively demonstrated in Section 6). Moreover, a fundamental deficiency of Eq. (2.2) is the dependence of the renormalizing item  $\bfGa(\bfx-\bfx_q)\lle\bfet\rle$ only on the average stress polarization tensor $\lle\bfet\rle$ while a corresponding item $\lle\bfGa(\bfx-\bfy)\bfet(\bfy)\rle_q(\bfx_q)$ ($\bfy\in \bfx_q$) in the new Eq. (4.1) explicitly depends on details distribution $\lle\bfet|v_q,\bfx_q\rle(\bfy)$ ($\bfy\in \bfx_q$). Because of this, the averaging methods used in Eq. (2.2) as a starting element conserve the mentioned deficiency of Eq. (2.2) (at least in some elements of these methods).
%\textcolor{blue}{
For example, Chen and Acrivos [36]
%(1978)
have estimated the effective elastic moduli through the accurate evaluation of binary interactions of inclusions without the hypothesis {\bf H1}, e.i. in our notations the operator $\bfcL_q^{\sigma}(\lle\overline{\bfsi}|;v_i,\bfx_i\rle_{q})(\bfx)$ (4.2) was estimated at the condition $\lle\widetilde{\bfsi}_{i,q}\rle\equiv \lle\bfsi\rle$ [compare with Eq. (2.17)].
However, the operator $\bfcL_q^{\sigma}(\lle\overline{\bfsi}\rle)(\bfx)$ (4.2) was estimated at the approximation (4.17) that implicitly implies the use of both the hypothesis {\bf H1} and Eq. (2.2) (see for details Subsection 10.2.2 in [6]).
%Buryachenko, 2007}).

On the other hand, although the method  (4.18)-(4.21) allows the inhomogeneous statistically averaged tensor $\lle\bfet\rle_{q}(\bfy)$
($\bfy\in v_q$) (4.21), but Eq. (4.21) containing the item $\lle\bfR^v\bfQ^0V\rle$ generated by the second summand in the left-hand side of Eq. (4.18) depending only on the average stress polarization tensor $\lle\bfet\rle$. As a consequence of this the final classical representations of the effective properties (2.19), (2.20)  and (2.25)
 depend only on average stress concentrator factors
 $\bfR_i$ and $\bfF_i$ while the effective properties (4.26) explicitly depend on the inhomogeneous tensors  $\bfR_i(\bfx)$ and $\bfF_i(\bfx)$ as well as on detailed distribution $\lle\bfet\rle_{q}(\bfy)$ ($\bfy\in v_q$) (4.23).

Moreover, the detected explicit dependence of the effective properties (4.26) on the detailed stress concentrator factors $\bfR_i(\bfx)$ and $\bfF_i(\bfx)$ rather than on the average values  $\bfR_i$ and $\bfF_i$ allows for an abandonment of the hypothesis {\bf H1b} [or (4.17)] whose accuracy is questionable for the inhomogeneous (e.g., coated) inclusions. In such a case the statistical average effective field  estimated by Eq. (4.22) is found to be inhomogeneous that discards the hypothesis {\bf H1a}. Quantitative estimations of the result of this abandonment of the hypothesis {\bf H1} will be performed in Section 6 in the framework of the hypothesis {\bf H3} for some particular cases of fiber composites.

\medskip
\noindent{\it 5.2 Analysis of the proposition 2 and hypotheses {\bf H1b} and {\bf H3}}
\medskip

As was mentioned, forfeiting of the effective filed hypothesis {\bf H1a} by the additional perturbation hypothesis (4.17) leads to Eq. (4.18) providing homogeneity of the effective field estimation for the ellipsoidal excluded volume $v_i^0$. Moreover this estimation of the effective field (and, therefore, of effective moduli) is invariant with respect to the size of the ellipsoidal excluded volume $v_i^0$. However, the additional hypothesis (4.17) is exactly fulfilled  only for the homogeneous ellipsoidal inhomogeneity $v_i$. For both the inhomogeneous and nonellipsoidal inclusions the equality (4.17) is just an approximation and the new general equation (4.1) has an advantage with respect to the popular one (2.2) only based on average strain polarization $\bfet_q$ [rather than $\lle\bfGa(\bfx-\bfy)\bfeta(\bfy)\rle_q$]. Then the size of the excluded volume $v_i^0$ will impact on the effective field  (4.22). Indeed, if the radius of the excluded volume $v_i^0$ in Fig. 1 increases  from $2a$ to $3a$ then the long distance of influence zone of the inhomogeneity $v_q$ on the effective field $\lle\overline{\bfsi}\rle_{i}(\bfx)$ will increase from the value $|\bfx-\bfx_q|=3a$ (as in Fig. 1) till $|\bfx-\bfx_q|=4a$. This influence will be quantitatively estimated in the next section.

A popular explanation of acceptance of the ``ellipsoidal symmetry" hypothesis (2.24)
is that this hypothesis just simplifies Eq. (2.23) reducing this equating to Eq.  (2.25) which does not contain the integrals. In a similar manner, a destination of the assumption of the ellipsoidal shape of the excluded volume $v_i^0$ in the hypothesis {\bf H3} is that this hypothesis just simplifies Eq. (4.18) by the use of analytical known tensor $\bfQ_i^0$ (expressed through the Eshelby tensor $\bfS_i^0$ (3.4)) which is exploited instead of a general tensor $\bfQ_i^0(\bfx)$ found numerically (see e.g. Subsection 4.7.4 in the book
[6]).
%by Buryachenko (2007)].
However, the both mentioned assumptions of the hypothesis {\bf H3} have a fundamental conceptual sense rather than only an analytical solution of some particular problem. Exploiting the Eshelby tensor concept in Eq. (4.18) (and in the MEFM) is based on the ellipsoidal shape of the correlation hole $v^0_{i}$ rather than
on the inclusion shape $v_i$. An abandonment of either the assumption of the $v^0_{i}$'s ellipsoidal shape or ``ellipsoidal symmetry" hypothesis (2.24) with necessarily leads to the inhomogeneity of the effective field $\overline{\bfsi}_i$ acting on the inclusion $v_i$ that is prohibited for the classical version of the MEFM. However, Buryachenko [6] %(2007)
(see Section 9.4) proposed a method for solution of Eq. (4.18) based on the general integral equation (2.2). Namely,
acting on Eq.  (4.18) by the operator $\bfcB=(\bfI-{\bf \Gamma}\bfM_1)^{-1}V_i$
yields

\vspace{-.5cm}
\noindent
\hspace{5.1cm}\begin{minipage}[b]{0.29\linewidth}
\centering
\epsfig{figure=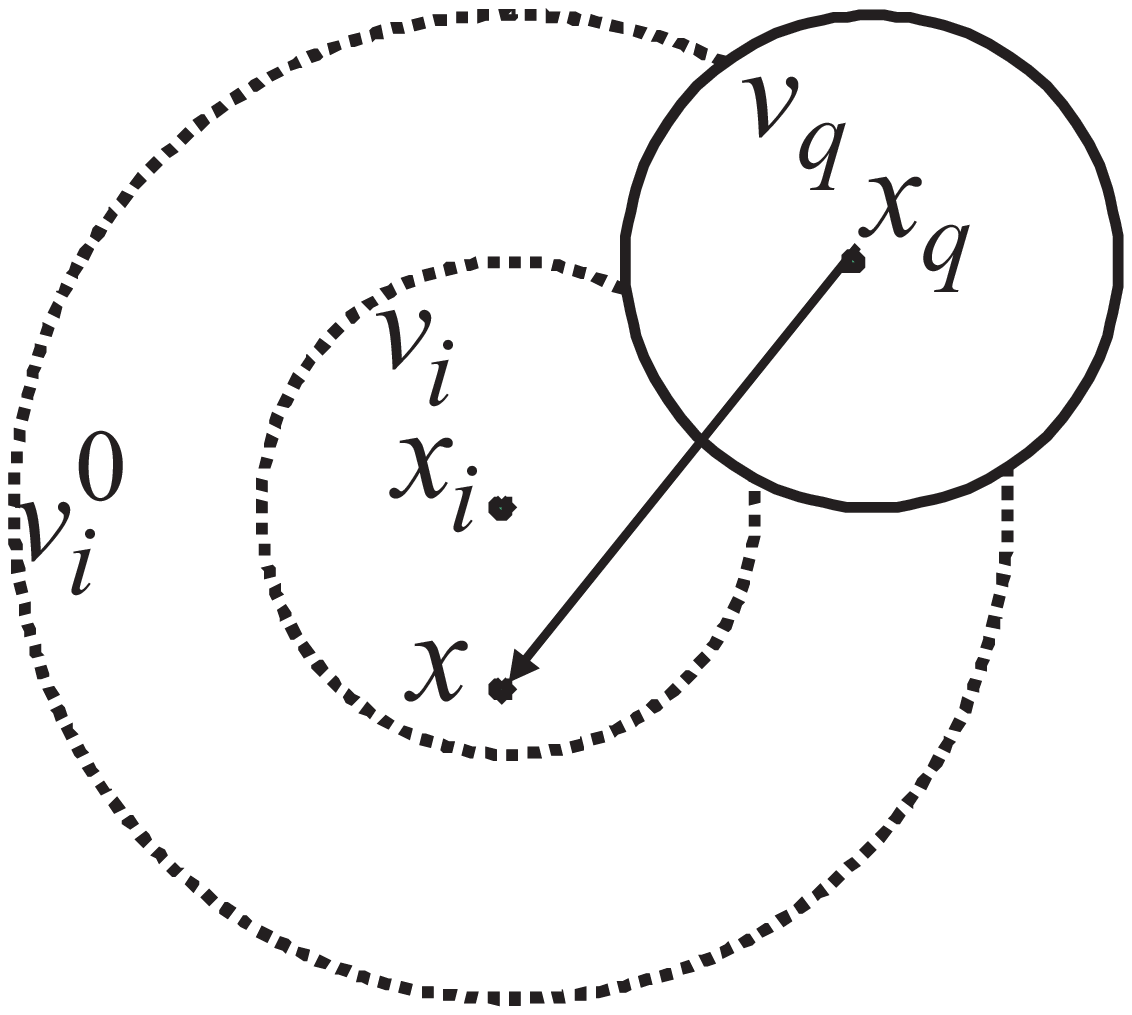,width=\linewidth}\\
\vspace{0.0cm}

{\sc Fig. 1}: {\tenrm
 Schematic mutual placements of $v_q,v_i$ and $v_i^0$
%}                   \end{figure}
}\end{minipage}%\hfill

\BB
\lle\bfet\rle_i(\bfx)=\bfR_i(\bfx)\lle\bfsi\rle+\bfF_i(\bfx)+\bfR^{{\bf Q}0}(\bfx)\sum^N_{q=1} \langle {\bfeta}_q \rangle c^{(q)}.  %(5.1)
\EE
where the numerical estimation scheme  of the tensor $\bfR^{{\bf Q}0}(\bfx)$ for any shape of $v_i^0$ was considered in Subsection 9.4.4 in Ref. [6]. %by Buryachenko (2007).
Volume averaging of Eq. (5.1) over the heterogeneity $v_i$ and summation over the inclusion number $i$ lead to the average strain polarization tensor $\lle\bfet\rle$ and, therefore,
 gives the representations for the effective properties
\BBEQ
\bfM^* &=& \bfM^{(0)}+
\bigl[\bfI-\lle \bfR^{{\bf Q}0v}V\rle\bigr]
^{-1}\lle \bfR^v V\rle,\\ %(5.2)
\bfbe^* &=&  \bfbe^{(0)}+
\bigl[{\bf I}-\lle \bfR^{{\bf Q}0v}V \rle\bigr]
^{-1} \lle {\bfF}^v V\rle. %(5.3)
\EEEQ
If (and only if) the correlation hole $v_i^0$ is chosen as an ellipsoid (homothetical, for example, to $v_i$) then the tensor ${\bfQ}^0_i(\bfx)=\bfQ_i^0\equiv{\rm const}$.,
$\bfR^{{\bf Q}0}_i(\bfx)=\bar v_i(\bar v_i^0)^{-1}\bfR_i(\bfx)\bfQ_i^0$,
$\ \bfB^{{\bf Q}0}_i(\bfx)=\bfB^0_i(\bfx)\bfQ_i^0$, and, therefore,
Eqs.  (5.2) and (5.3) are reduced to the known representations (2.20) and (2.25) with the constant tensor $\bfQ^0_i$ depending on the orientation of the correlation hole $v_i^0$.
An advantage of Eqs. (4.22)--(4.26) with respect to Eqs. (5.2) and (5.3) are that Eqs. (5.2) and (5.3) are fundamentally limited by analysis of statistically homogeneous media subjected to the homogeneous boundary conditions while the system (4.22) and (4.23) can be easily generalized to the statistically inhomogeneous media. Moreover, the method (5.1)-(5.3) conserves a fundamental deficiency of the general integral Eq. (2.2) containing the renormalizing item defining only by the averaged strain polarization tensor $\lle\bfet\rle$. Estimation of the tensor  $\bfR^{{\bf Q}0}_i(\bfx)$ in an auxiliary model problem with homogeneous fictitious eigenstrain in $v_i^0$ implies that influence of surrounding inclusions $v_q$ ($\bfx_q\in v_i^0$) is defined by the average strain polarization tensors $\lle\bfet\rle_q$ rather than its detailed distribution $\lle\bfet\rle_q(\bfx)$ ($\bfx\in v_q$). Impact of the last improvement on the estimated effective properties will be considered in Section 6 for the circle $v_i^0$ although, of course, analysis of any shape of $v_i^0$ present no additional difficulties for the new method (4.22)-(4.23) as opposed to the method (5.1)-(5.3) requiring evaluation of the supplementary tensor $\bfR^{{\bf Q}0}_i(\bfx)$.

\medskip
\noindent{\it 5.3 Analyses of the hypotheses {\bf H2a} and {\bf H2b}}
\medskip

As it was noted in Subsection 2.3, the hypotheses {\bf H2a} and {\bf H2b} are not conceptually dependent on the hypothesis {\bf H1} and can be applied in general case even if the hypothesis {\bf H1} is violated. Indeed, at obtaining of Eq. (2.18) we already used the hypothesis {\bf H1} because
Eq. (2.18) contains the objects $\bfR_q\lle\overline{\bfsi}\rle_q+\bfF_q$ instead of their
operator generalization $\bfcL^{\eta}_q(\bfeta)(\bfx)$ (3.18) which does not use the hypothesis {\bf H1}.  However, even in this case the effective field
$\lle\overline{\bfsi}\rle_i(\bfx)$ (2.18) is an inhomogeneous function of the coordinate $\bfx\in v_i$. In actual truss, a subsequent averaging of Eq. (2.18) over the inclusion $v_i$ is tantamount to a secondary using of the hypothesis {\bf H1} that is not necessary and can be avoided. However, such an inhomogeneity of $\lle\overline{\bfsi}\rle_i(\bfx)$ (2.18)
is beyond the scope of the current study and will be analyzed in other publications. It is correctly noted that application to Eq. (2.18) of the simplified hypothesis {\bf H2b} [with $\bfZ_{ij}=\bfI\delta_{ij}$ (2.22)] in an accompany with the hypothesis  {\bf H3} does not fall outside the scope the hypothesis {\bf H1} and reduces Eq. (2.18) to Eq. (4.18) with subsequent obtaining of the known representations for the effective properties (2.20) and (2.25). However, the eventual abandonment of the hypothesis {\bf H1} can be done before the use of the hypotheses either {\bf H2a} or {\bf H2b} as it was performed in Eq. (4.2). Then the following solution of Eq. (4.2) by the use of the hypotheses
either {\bf H2a}, {\bf H2b} or {\bf H3} does not lead to the necessity of using the hypothesis {\bf H1} that will be quantatively demonstrated in Section 6 at some numerical examples.

\bigskip
\noindent{\bf 6. Numerical results}
\medskip

\setcounter{equation}{0}
\renewcommand{\theequation}{6.\arabic{equation}}

With the non-essential restriction on space dimensionality $d$
and the shape of inhomogeneities we will consider 2-$D$ problems for composites reinforced by cylindrical infinite fibers.
The domains of inclusions $v_i$
are discretized along the polar angle and the radius in the local polar coordinate
system with the centers $\bfx_i$.
Then the points
\BB
\Big\{(r,\varphi)\ |\ (p-1){2\pi\over l}<\varphi<p{2\pi \over l},\
(q-1){a_i\over m}<r<q{a_i\over m}\Big\} %(6.1)
\EE
($p=1,2\ldots,l;\ q=1,2,\ldots,m$) represent  the elements of $\Gamma_i^{pq}$
of the meshes $\Omega_i$ ($i=2,\ldots,n$)
that is not optimized, but is efficient.
Moreover, the square meshes
\BB
\Big\{(x_1,x_2)^{\top}\ |\ (p-1){a_i\over l}<x_1<p{a_i \over l},\
(q-1){a_i\over l}<x_2<q{a_i\over l}\Big\}, %(6.2)
\EE
where $x_1,\ x_2$ are local coordinates with origins at the fiber centers,
will be used for stress estimation inside and outside the fiber.
We will use piecewise-constant elements of the meshes which are
not very cost-efficient but are very easy  for computer programming,
and the discretization (6.2) permits the analysis of nonregular
inclusion shapes.  For simplicity estimation of integrals
involved we will utilize  the basic numerical integrations formulas of  Simpson's
rule and trapezoidal rule for the uniform (6.1)-(6.2) and nonuniform meshes
considered below, respectively.

We detected that in the concrete examples
of high matrix-inclusion elastic contrast considered and  some others,
the standard popular iterative schemes (3.14)
may diverge or converge very slowly (i.e. the iteration scheme (3.12) does not work in general)
so that an implementation of the { improved}
algorithm { proposed in this paper} becomes more complicated.
In such a case, following Refs. [37], [38],
%Perlin (1976) and Mikhlin and Pro\"ossdorf (1980),
we introduced the subsidiary grid of the support points $\bfze_j$
at the centers of each elements additionally to the nodal points $\bfs_j$ at the apexes of elements. After determining in this way at all the points $\bfze_j$ the values
of the function $\bfet^{[1]}(\bfze_j)$,
we find its values at the nodal points $\bfet^{[1]}(\bfs_j)$
by linear interpolation, and so on (see details in Ref. [38]).
%Mikhlin and Pro\"ossdorf, 1980).
Moreover,
instead of the point Jacobi iteration method
displayed in Eq. (3.14)
we use the accelerated Liebmann method (called also extrapolated Gauss-Seidel method)
which is usually ``faster" than the point Jacobi method,
and has the computational advantage that it does not require the
simultaneous storage of the two iterations $\bfet_{(k+1)}$ and
$\bfet_{(k)}$ (see, e.g., [39]). % Varga, 2000).
The convergence of the
scheme (3.14) is provided by their modification
$ \bfet^{[k+1]}={1\over2}[\bfet^{[0]}+ \bfet^{[k]}+\bfcK_i\bfet^{[k]}] $
(see for details Ref. [38]). %Mikhlin and Pro\"ossdorf, 1980).
It should be mentioned that in forthcoming numerical examples
we will use only the iteration scheme described above. Comparative analysis
of this scheme with other known iteration schemes is beyond the scope of the current paper.
Moreover, although the convergence of this method was rigorously proved
in Ref. [38] %by Mikhlin and Pro\"ossdorf (1980)
for the elastic problems of an arbitrary
dimensions, we  will demonstrate its effectiveness only for 2-D problems;
the analysis of 3-D problems is beyond the scope of this paper.

We consider a pure mechanical problem ($\bfbe\equiv {\bf 0}$) and assume the matrix is epoxy resin
($\bfL^{(0)}=(3k^{(0)},2\mu^{(0)})$, $k^{(0)}=3.83$ GPa and $\mu^{(0)}=1.27$
GPa) which contains identical circular glass fibers
($\bfL^{(1)}=(3k^{(1)},2\mu^{(1)}),\ \  k^{(1)}=34.3$ GPa and $\mu^{(1)}=31.3$ GPa).
If the pair distribution
  function $g({\bf x}_i-{\bf x}_m)\equiv
  \varphi(v_i, {\bf x}_i\vert; v_m,{\bf x}_m)/n^{(k)}$
  depends on $\vert {\bf x}_m-{\bf x}_i\vert$  it is called the
  radial distribution function (RDF).  Two alternative RDFs of inclusion will be examined (see Refs. [40], [41])  % Torquato and Lado, 1992; Hansen and McDonald, 1986)
\BBEQ
\!\!\!\! g({\bf x}_i-{\bf x}_q)\!\!\!\!&\equiv &\!\!\!\!
\varphi(v_i, {\bf x}_i\vert; v_q,{\bf x}_q)/n^{(q)}
=H(r-2a),\\
%\eqno (6.3)$$
\!\!\!\!g({\bf x}_i-{\bf x}_j) \!\!\!\!&=&\!\!\!\! H(r-2a)
%\nonumber\\\!\!\!\! &\cdot & \!\!\!\!
\bigg\{1+{4c \over \pi}\Big[ \pi-2\sin^{-1}({r\over 4a})-{r\over 2a}\sqrt
{1-{r^2\over 16a^2}} \Big]H(4a-r) \bigg\}
%\eqno (6.4)$$
\EEEQ
where $H$ denotes the Heaviside step function, $r\equiv \vert
{\bf x}_i-{\bf x}_q\vert $ is the
distance between the nonintersecting inclusions $v_i$  and $v_q$,
and $c$ is the volume fraction of fibers of the radius $a$.
The formula (6.4) takes into account a neighboring order in the distribution of the inclusions.

At first we will perform our evaluations for composites with homogeneous fibers  described by the RDF (6.3) in the framework of the hypotheses {\bf H2b} and {\bf H3}. Influence of the effective field hypothesis {\bf H1} is considered by comparison of statistical averages of stresses in the fibers  estimated by the classical approach $\lle\bfsi\rle_i^{\rm old}(\bfx)\equiv$const. (4.15), (4.18) as well as by the proposed one
$\lle\bfsi\rle_i^{\rm new}(\bfx)$ (4.22), (4.23). We considered a volume fraction of fibers $c=0.65$  and evaluated the stress perturbations
$\bfcL^{\sigma}_q(\lle\bfet\rle_q^{[n]})(\bfx)$ (4.22) in the vicinity
$\{\bfx|\, {\max}_{x}|\bfx-\bfx_q|=3a\}$
of the area $v_q$ rather than only a stress distributions in the inhomogeneity $\bfx\in v_q$.
Then $\lle\bfsi\rle_i^{\rm old}$ and $\lle\bfsi\rle_i^{\rm new}(\bfx)$ differ from one another no more than $0.09\%$ that coincides with a computational error realized in the method (4.22), (4.23)
for two different meshes (6.2) with $l=15$ and $l=30$. Thus, we qualitatively proved that in the considered example both methods the old  (4.15), (4.18) and new (4.22), (4.23) ones which are based on the classical (2.2) and new (4.1) general integral equations, respectively,  lead to the same numerical results. This conclusion quantitatively confirms the Proposition 1) establishing an equivalentness of Eqs. (2.2) and (4.1) for statistically homogeneous fields of homogeneous ellipsoidal heterogeneities subjected to the homogeneous boundary conditions. Now we will consider an influence of incorrect using of the operator $\bfcL^{\sigma}_q(\lle\bfet\rle_q^{[n]})(\bfx)$ (4.22) when only $\bfx\in v_q$ are considered, which means that stress perturbations introduced by the moving inhomogeneity $v_q$ in their vicinity $\{\bfx|\, a<|\bfx-\bfx_q|<3a\}$
are neglected. For this purpose, the means of stress concentrator factors
($\bfbe\equiv{\bf 0}$)
\BB
\lle\bfsi\rle_i(\bfx)=\bfB^*(\bfx)\lle\bfsi\rle, %(6.5)
\EE
defined analogously to Eq. (4.25), will be estimated. In Fig. 2 the components $B^*_{2211}(\bfx)$ demonstrating maximum dependence on  $\bfx=(x_1,0)^{\top}$ are presented for the initial  [$B^{*[0]}_{2211}(\bfx)=B^{*{\rm old}}_{2211}\equiv$const.], second [$B^{*[2]}_{2211}(\bfx)$], forth [$B^{*[4]}_{2211}(\bfx)$], and tenth [$B^{*[10]}_{2211}(\bfx)$] iterations of stress concentrator factor. A fast convergence of the proposed iteration method can be seen: the tenth iteration differs from the ninth, fourth, and initial approximations by 0.013\%, 0.084\%, and 2.9\%, respectively. In so doing, the difference 2.9\% essentially exceeds the possible errors of both the calculations and iteration scheme.
Thus, for statistically homogeneous fields of homogeneous circle inclusions subjected to the homogeneous boundary conditions, the old and new approaches based on the backgrounds in the form of Eqs. (2.2) and (4.1), respectively, lead to equivalent results. Thus, in the
case of the background (4.1) we must estimate the stress perturbation in the vicinity $\{\bfx|\, a<|\bfx-\bfx_q|<3a\}$ of the moving inhomogeneity $v_q$. This statement contradicts to the proposition 1 obtained at the use of the old background (2.2).

\vspace{-0.1cm}
\noindent
\hspace{3.65cm}\begin{minipage}[b]{0.4\linewidth}
%\hspace{1.65cm}\begin{minipage}[b]{0.75\linewidth}
\centering
\epsfig{figure=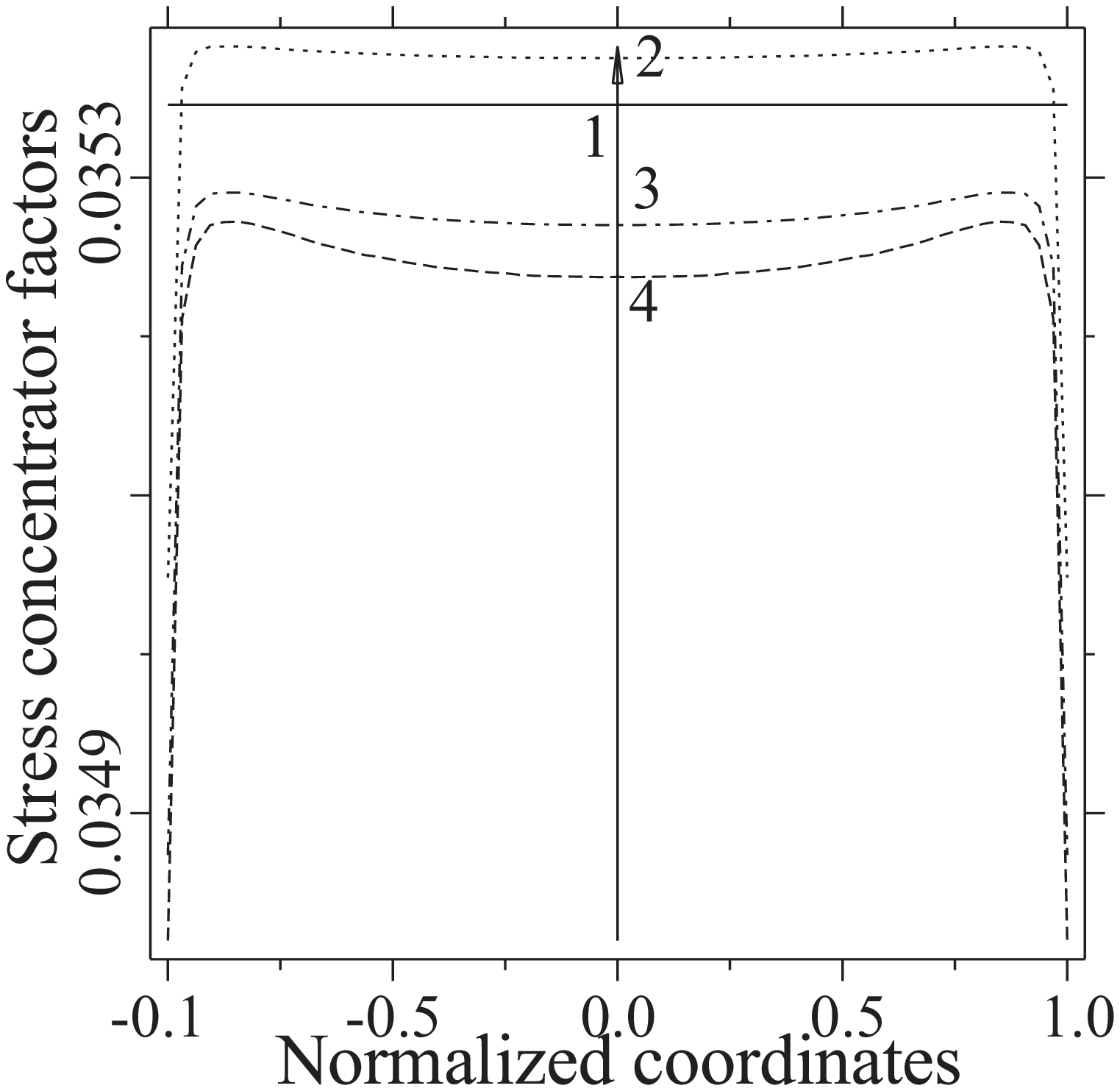,width=\linewidth}\\
\vspace{-0.0cm}

{\sc Fig. 2}:{\tenrm  $B^{*[k]}_{2211}(x_1)$ vs
$x_1/a$ in homogeneous fiber: curves 1, 2, 3, 4 for $k=0, 2,4,12$, respectively
%}                   \end{figure}
}\end{minipage}%\hfill

%\noindent

  We are expected to get a larger difference of the backgrounds (2.2) and (4.1) for composites reinforced by either nonellipsoidal or inhomogeneous inclusions demonstrating essentially inhomogeneous stress distribution inside inclusions even in the framework of the hypothesis {\bf H1}. The interphase is usually the product of processing conditions involved in composite manufacture. In relation to this problem, an application of the concept of functionally graded materials by Hirai {\it et al.} [42]
%(1987)
for description of the interphase whose moduli may vary continuously is worthy of notice. Along
this line one may, for instance, refer to the works [43-47]
% of Theocaris (1987),  Jayaraman and
% Reifsnider (1992), Wang and Jasiuk (1998), Weng (2003), You {\it et al.} (2006)
concerned with the spatially nonuniform properties of interphase.
Just for concreteness, we assume that fibers contain the cores of the radius $a^c<a$ with the constant moduli $\bfL^{(1)}\equiv$const while the moduli $\bfL^{\rm int}(\bfx)$ in the interphase with the coating thickness $h=a-a^c$
are taken to vary linearly with the radial distance $r=|\bfx|:$
\BB
\bfL^{\rm int}(r)=\bfL^{(0)}+(\bfL^{(1)}-\bfL^{(0)})(a-r)/h. %(6.6)
\EE
For demonstration of maximum difference between the old and new approaches, we will consider in detail a thick coating with the relative coating thickness $h/a^c=0.5$ although other ratios $h/a^c$ will be also analyzed in a few comparative examples. At first, we will analyze results obtained in the framework the hypotheses {\bf H2b} and {\bf H3} for the RDF (6.3).
In Fig. 3 the iterations $B^{*[k]}_{1111}(\bfx)$ ($k=0,2,4,12$) at the axis $\bfx=(x_1,0)^{\top}$ are presented for $c\equiv \pi a^2n=0.65$.  The initial approximation $B^{*[0]}_{1111}(\bfx)$ corresponding to the classical estimation (4.20) and using the old background (2.2)  reveals their essential inhomogeneity ($14\%$) even in the framework of the effective field hypothesis {\bf H1}.
The new background (4.1) allow the use of this inhomogeneity for refinement of the renormalizing item in Eq. (4.14) without exploiting of the hypothesis {\bf H1}. The twelfth iteration $B^{*[12]}_{1111}(\bfx)$ differs from the initial approximation $B^{*[0]}_{1111} (\bfx)$ by 10.7\%  while the 12th and 11th iterations are distinguished from one another by
0.9\%. Of even greater difference of results obtained for the backgrounds (2.2) and (4.1) is observed for the component $B^{*[k]}_{1122}(\bfx)$ ($k=0,2,4,12$) at $\bfx=(x_1,0)^{\top}$
in Fig. 4. Indeed, $B^{*[0]}_{1122}(\bfx)>0$ at any $\bfx=(x_1,0)^{\top}$ while $B^{*[k]}_{1122}(\bfx)<0$  at $-0.75a<x_1<0.75a$. Again, the proposed iteration method converges rapidly and we contend that 12th iteration provides a difference from 11th iteration of 0.2\%.

\vspace{-1.0mm}
\noindent
\begin{minipage}[b]{.42\linewidth}
\centering\epsfig{figure=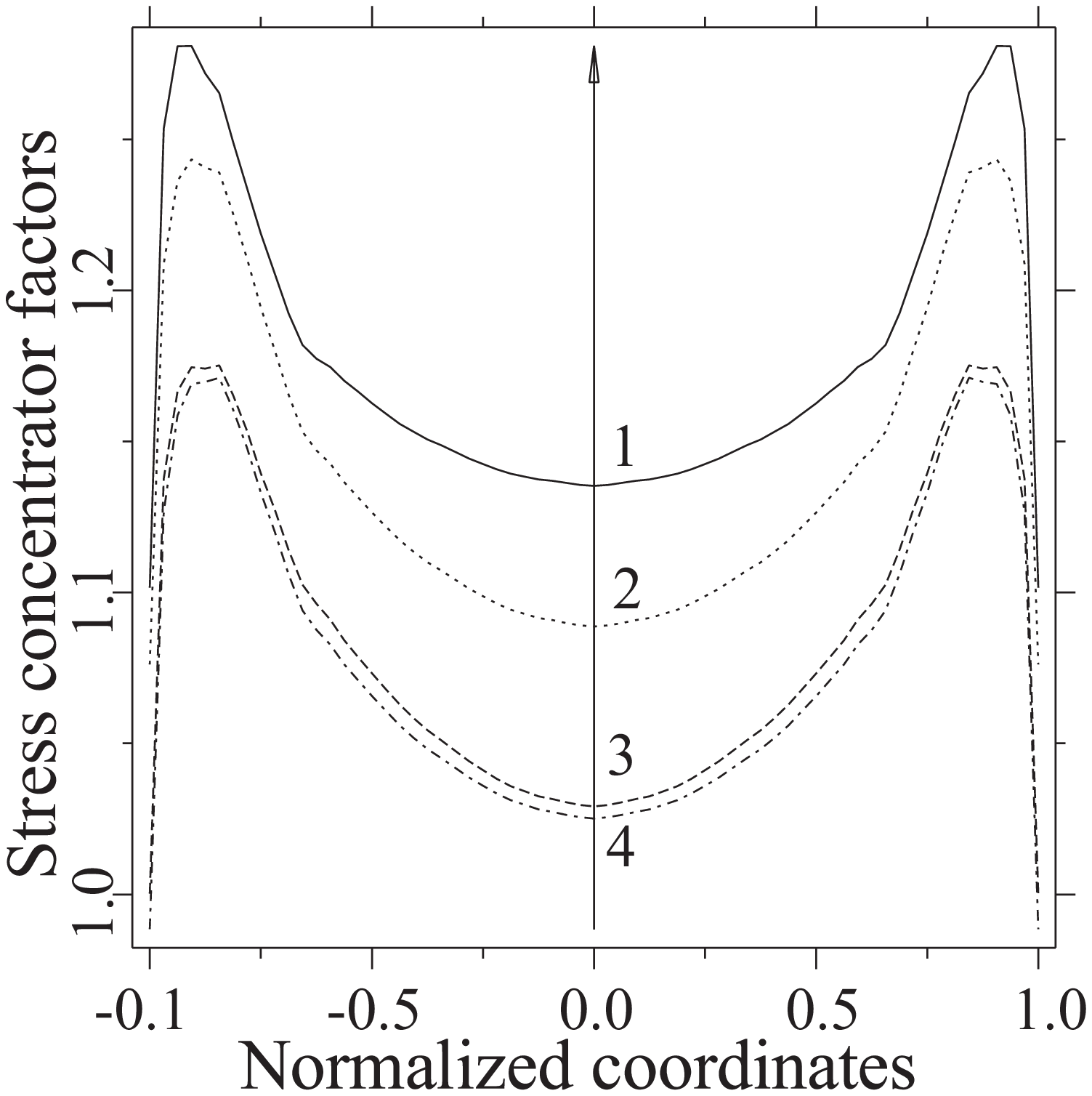,width=\linewidth}\\
\vspace{-0.0cm}
\tenrm {\sc Fig.3}:
%Stress concentrator factors
$B^{*[k]}_{1111}(x_1)$ vs
$x_1/a$ in inhomogeneous fiber: curves 1, 2, 3, 4 for $k=0, 2,4,12$, respectively.
%\hspace{0.5cm}
\end{minipage}\hfill
\begin{minipage}[b]{.42\linewidth}
\centering\epsfig{figure=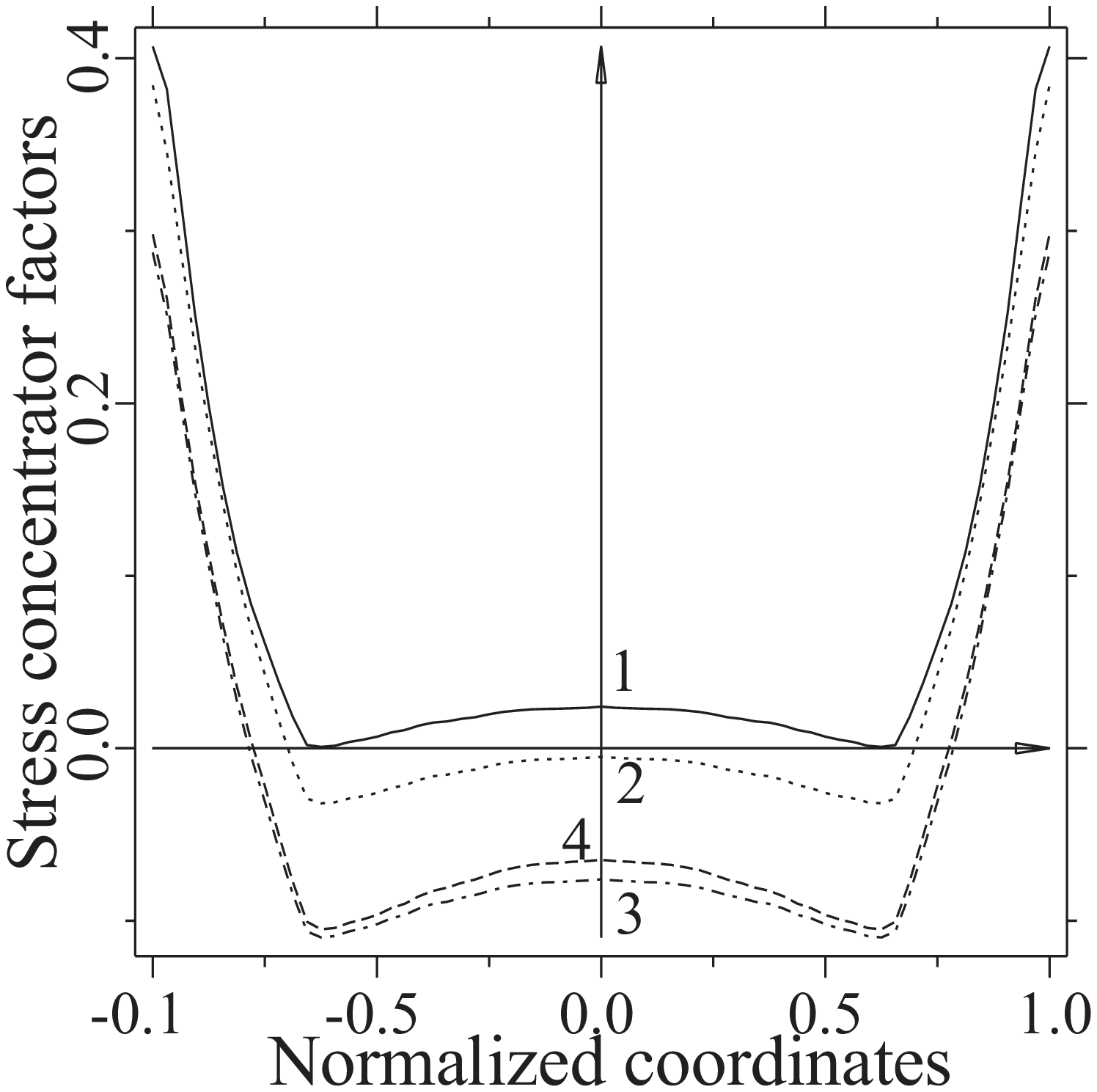,width=\linewidth}\\
\vspace{-0.0cm}
\tenrm {\sc Fig.4}:
%Stress concentrator factors
$B^{*[k]}_{1122}(x_1)$ vs
$x_1/a$ in inhomogeneous fiber: curves 1, 2, 3, 4 for $k=0,2,4,12$, respectively.
\end{minipage}

So much prominent and systematic differences of the old and new approaches are based on the abandonment from effective field hypothesis {\bf H1} in the new approach.
We estimated a tensor of effective stress concentrator factor
\BB
\lle\overline{\bfsi}\rle(\bfx)=\overline{\bfB}^*(\bfx)\lle\bfsi\rle, \ \
(\bfx=(x_1,0)^{\top}) %(6.7)
\EE
 and presented the components of their $k$-th approximations $\overline {B}^{*[k]}_{1111}(\bfx)$ and $\overline {B}^{*[k]}_{1122}(\bfx)$ ($k=0,2,4,12$) in Figs. 5 and 6, respectively. $\overline {B}^{*[12]}_{1111}(\bfx)$ differs from both the classical
$\overline {B}^{*[0]}_{1111}(\bfx)$ (4.15), (6.7) and $\overline {B}^{*[11]}_{1111}(\bfx)$ on 8.1\% and 0.03\%, respectively, while $\overline {B}^{*[12]}_{1111}(\bfx)$ varies along $x_1$ over 2.2\%. However, we can observe
in Fig. 6 a significantly more dramatic situation with the component $\overline {B}^{*[k]}_{1122}(\bfx)$ where all iterations differ by a sign from  the classical one $\overline {B}^{*[0]}_{1122}(\bfx)$ (4.15), (6.7) almost at all values $|x_1|<a$.

We now turn our attention to the analysis of the size of the circle excluded volume $v_i^0$ with the radius $a^0$ on the stress concentrator factor ${\bfB}^{*[k]}(\bfx)$ also for the radial distribution function (6.3) reducing Eq. (4.22) to Eq. (4.24). We will compare the estimation of ${\bfB}^{*[k]}(\bfx)$ carried out for $a^0=3a$ with previously obtained results for $a^0=2a$ (see Figs. 3 and 4).
The components ${B}^{*[k]}_{1111}(\bfx)$ and  ${B}^{*[k]}_{1122}(\bfx)$ ($\bfx=(x_1,0)^{\top}$)
%\noindent
are presented in Figs. 7 and 8, respectively, for $k=0$ (curves 1) and
$k=12$ for
both $a^0=2a$ (curves
2) and $a^0=3a$ (curves 3). The RDF (6.3) provides the
``ellipsoidal symmetry" hypothesis {\bf H3} (2.24) and, because of this, the classical representations for ${\bfB}^{*[0]}(\bfx)$ is invariant to the size of $v^0_i$ while
${B}^{*[12]}_{1111}(\bfx)$ and ${B}^{*[12]}_{1122}(\bfx)$ estimated by the new approach
(4.24) for $a^0=2a$ and $a^0=3a$ differ at $x_1=0$ one from another by 3.1\% and 50\%, respectively.  Finally, we compare the influence of the RDF (6.3) and (6.4) at $a^0=2a$ on estimation of ${\bfB}^{*[k]}(\bfx)$. Needless to mention that ${\bfB}^{*[0]}(\bfx)$ (4.20) is invariant with respect to the RDF while ${B}^{*[12]}_{1111}(\bfx)$ and ${B}^{*[12]}_{1122}(\bfx)$ estimated for the RDF (6.3) (curve 2) and (6.4) (curves 4) are distinguished by 3.7\% and 33\%, respectively. The indicated differences demonstrating fundamentally new effects inherent in the new approach (4.22)-(4.25)
far exceed the iteration error between 11th and 12th iterations
%of corresponded components
which are less than 0.03\%.

\vspace{-0.0mm}
\noindent
\begin{minipage}[b]{.42\linewidth}
\centering\epsfig{figure=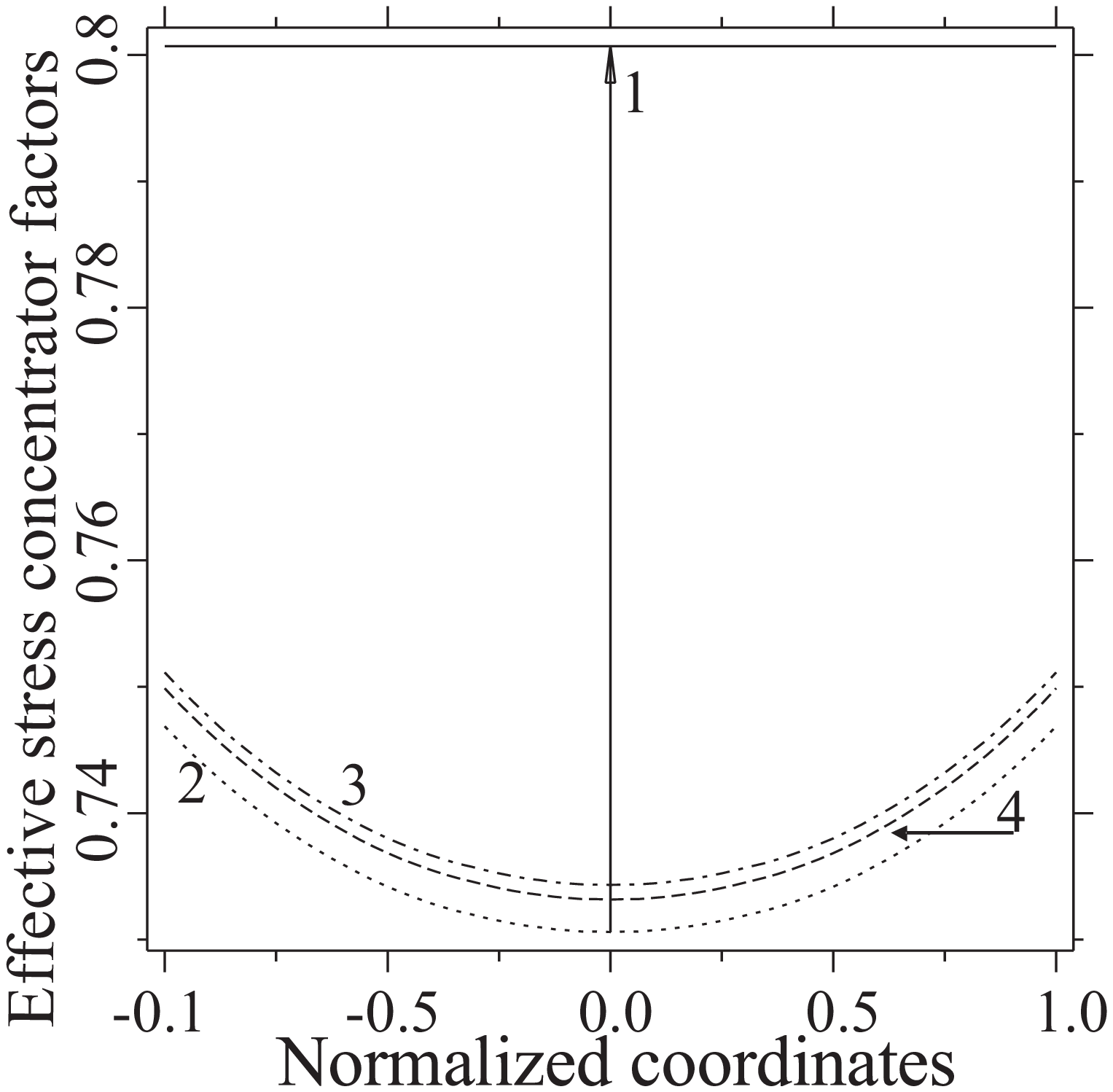,width=\linewidth}\\
\vspace{-0.0cm}
\tenrm {\sc Fig. 5}: Effective field concentrator factors ${\small \overline{B}^{*[k]}_{1111}(x_1)}$ vs
$x_1/a$ in inhomogeneous fiber: curves 1, 2, 3, 4 for $k=0,2,4,12$, respectively.
%\hspace{1cm}
\end{minipage}\hfill
\begin{minipage}[b]{.42\linewidth}
\centering\epsfig{figure=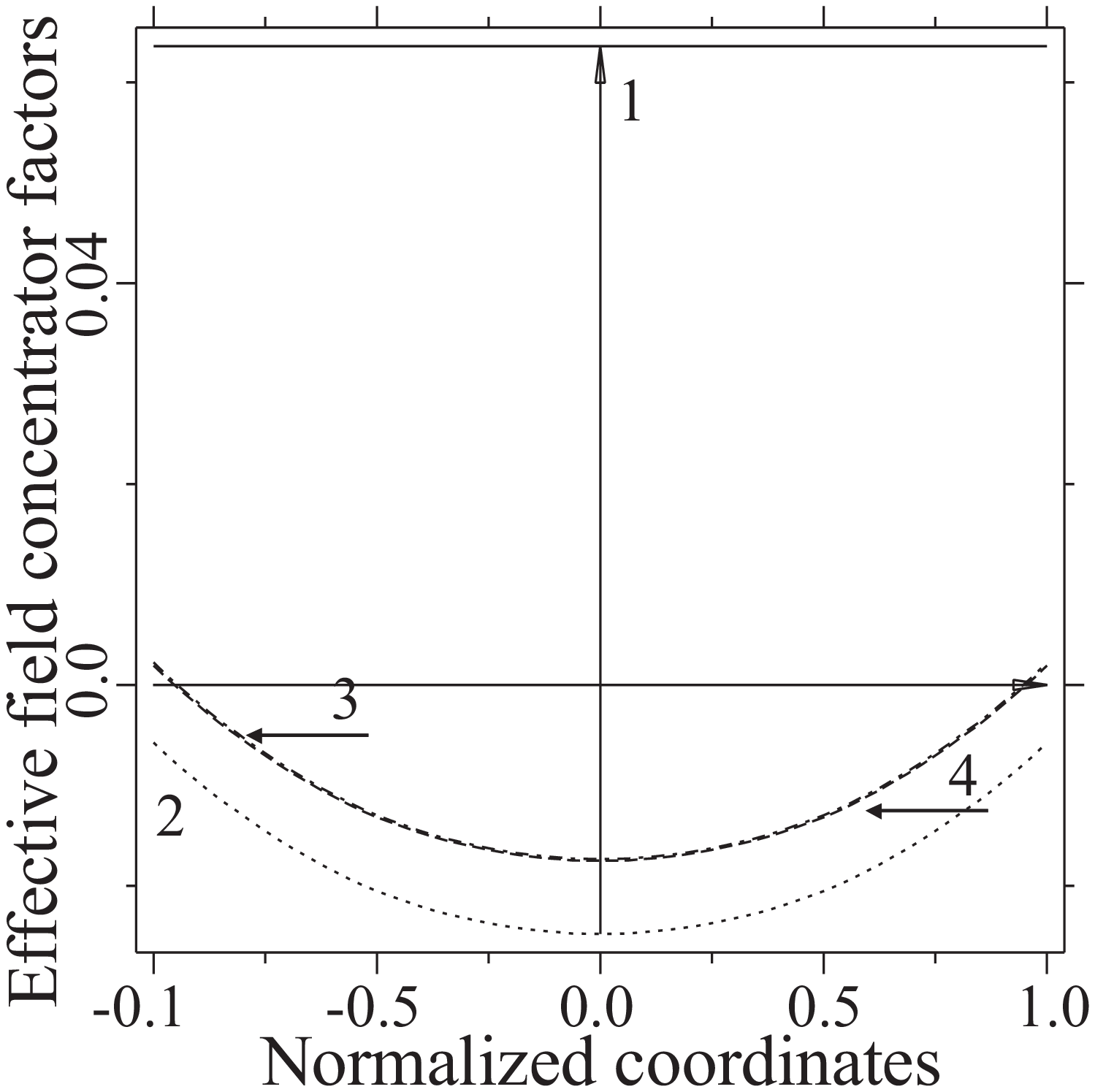,width=\linewidth}\\
\vspace{-0.0cm}
\tenrm {\sc Fig. 6}: Effective field concentrator factors ${\small \overline{B}^{*[k]}_{1122}(x_1)}$ vs
$x_1/a$ in inhomogeneous fiber: curves 1, 2, 3, 4 for $k=0,2,4,12$, respectively
\end{minipage}

\vspace{1.0mm}
\noindent
\begin{minipage}[b]{.42\linewidth}
\centering\epsfig{figure=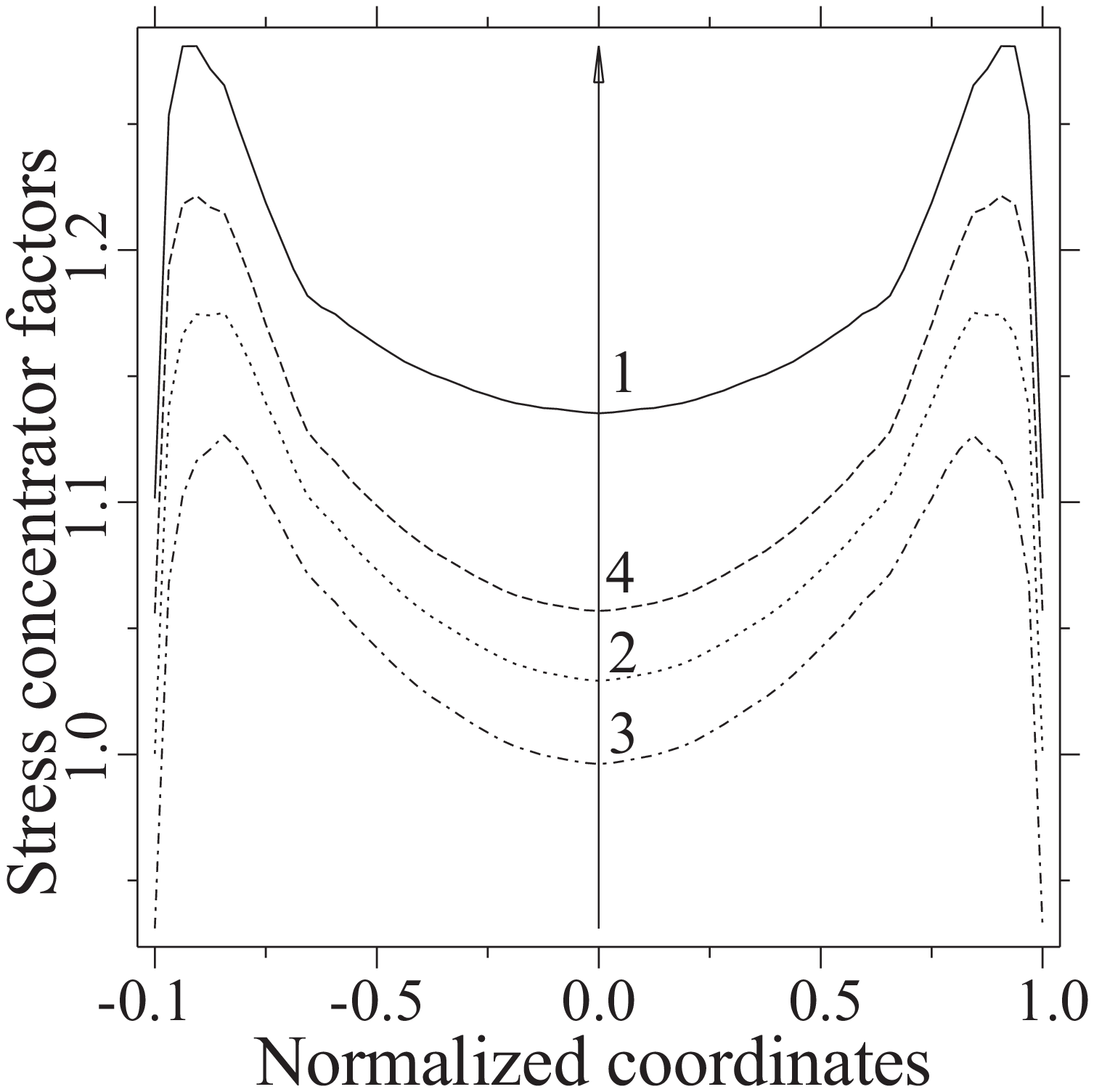,width=\linewidth}\\
\vspace{0.1cm}
\tenrm {\sc Fig. 7}:  $B^{*[k]}_{1111}(x_1)$ vs
$x_1/a$ for the different $a^0$ and RDF: curves 1 ($k=0$), 2
[RDF (6.3), $a^0=2a,\ k=12$], 3 [RDF (6.3), $a^0=3a,\ k=12$], 4
 [RDF (6.4), $a^0=2a, \ k=12$].
%\hspace{1cm}
\end{minipage}\hfill
\begin{minipage}[b]{.42\linewidth}
\centering\epsfig{figure=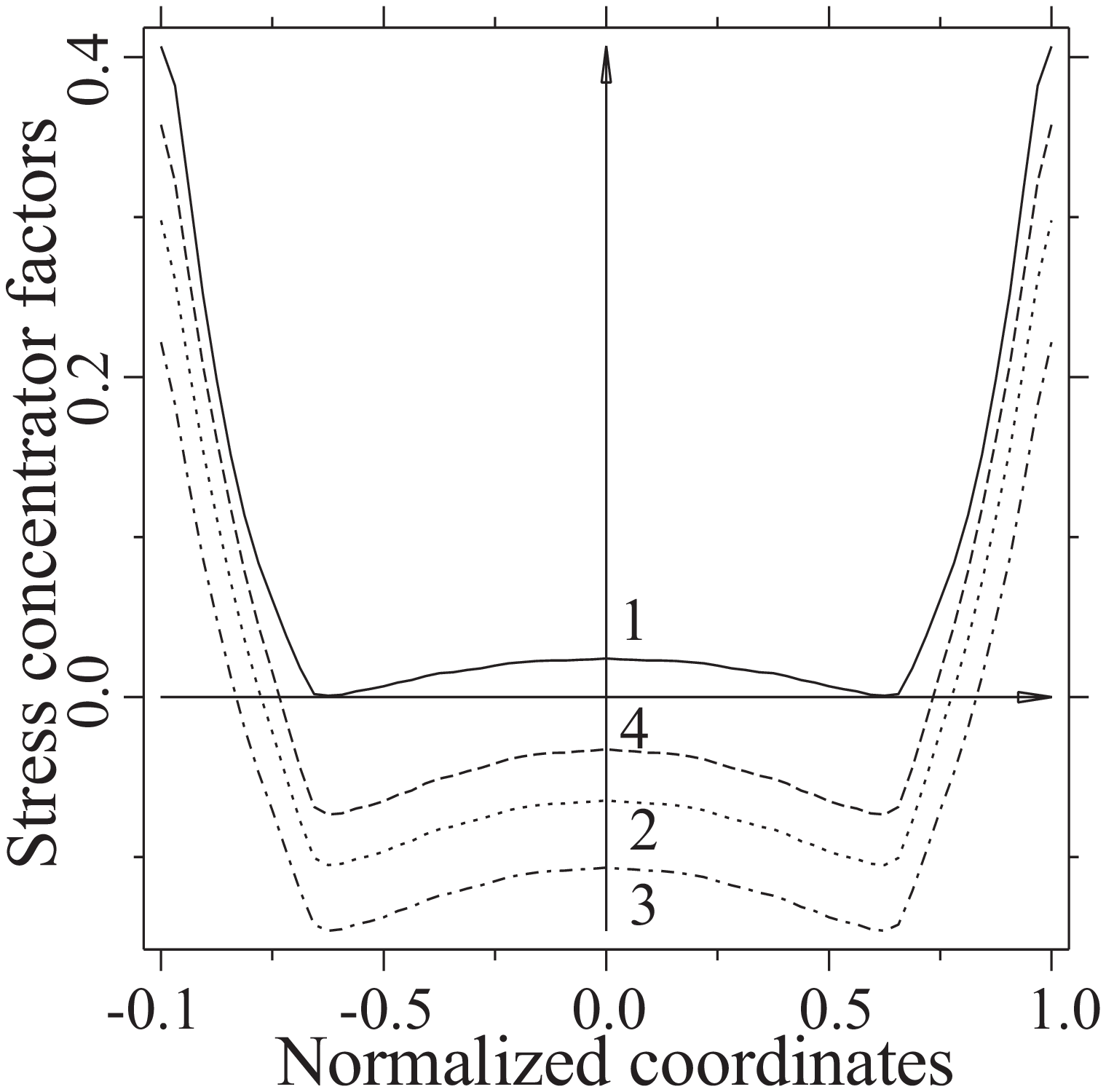,width=\linewidth}\\
\vspace{0.1cm}
\tenrm {\sc Fig. 8}: Fig. 8.  $B^{*[k]}_{1122}(x_1)$ vs
$x_1/a$ for the different $a^0$ and RDF: curves 1 ($k=0$), 2
[RDF (6.3), $a^0=2a,\ k=12$], 3 [RDF (6.3), $a^0=3a,\ k=12$], 4
 [RDF (6.4), $a^0=2a, \ k=12$]
\end{minipage}

Just for completeness, we will estimate an influence of the interphase thickness $h$ (6.6) on the stress concentrator factor ${\bfB}^{*[0]}(\bfx)$ and ${\bfB}^{*[12]}(\bfx)$ ($\bfx=(x_1,0)^{\top}$) for the RDF (6.3) with $a^0=2a$. In addition to Figs. 3 and 4 displaying the results for $h/a^c=0.5$, we are demonstrating the similar estimations ${B}^{*[k]}_{1122}(\bfx)$ ($k=0,12$) for $h/a^c=0.1,\ 0.25$, and $h/a^c=1$ in Fig. 9. As can be seen, ${B}_{1122}^{*[0]}(a)$ and ${B}_{1122}^{*[12]}(a)$ differ one from another by
20.5\%, 30.3\%, and  33.1\% for $h/a^c=0.1,\ 0.25$, and $h/a^c=1$, respectively.
The similar differences for the components ${B}_{1111}^{*[0]}(a)$ and ${B}_{1111}^{*[12]}(a)$ are 6.0\%, 7.2\%, and  8.5\%, respectively.

We complete our numerical analysis by estimation of isotropic effective moduli $\bfL^*=
2k^*_{[2]}\bfN_1+2\mu^*_{[2]}\bfN_2$ ($\bfN_1=\bfde\otimes\bfde/2,\ \bfN_2=\bfI-\bfN_1$).
For the fiber composites it is the plane-strain bulk modulus $k^{(0)}_{[2]}$
(and $k^*_{[2]}$) -- instead of the 3-D bulk modulus $k^{(0)}_{[3]}$ -- that plays the
significant role: $k^{(0)}_{[2]}=k^{(0)}_{[3]}+\mu^{(0)}_{[3]}/3$,
$\mu^{(0)}_{[2]}=\mu^{(0)}_{[3]}$.
%\noindent The normalized bulk and shear moduli $k^*/k^{(0)}$ and
$\mu^*/\mu^{(0)}$ are presented in Fig. 10 for both the
%\noindent
classical approach [corresponding to the stress concentrator factors $\bfB^{*[0]}(\bfx)$] and new one [corresponding to the 12th iteration $\bfB^{*[12]}(\bfx)]$. As can be seen, the distinctions between two approaches equal 3.8\% and 12.0\% for $c=0.65$ for $\mu^*/\mu^{(0)}$ and $k^*/k^{(0)}$, respectively.
In so doing, the stress concentrator factors in these approaches at the point $x_1=a$ of fibers can differ on 30\% and, moreover, these estimations for the different approaches can have the different signs at other same domains  of $v_1$ (see Fig. 4).  Thus, stress concentrator factors are significantly more sensitive values to the choice of the approach than effective elastic moduli.

\vspace{0.0mm}
\noindent
\begin{minipage}[b]{.42\linewidth}
\centering\epsfig{figure=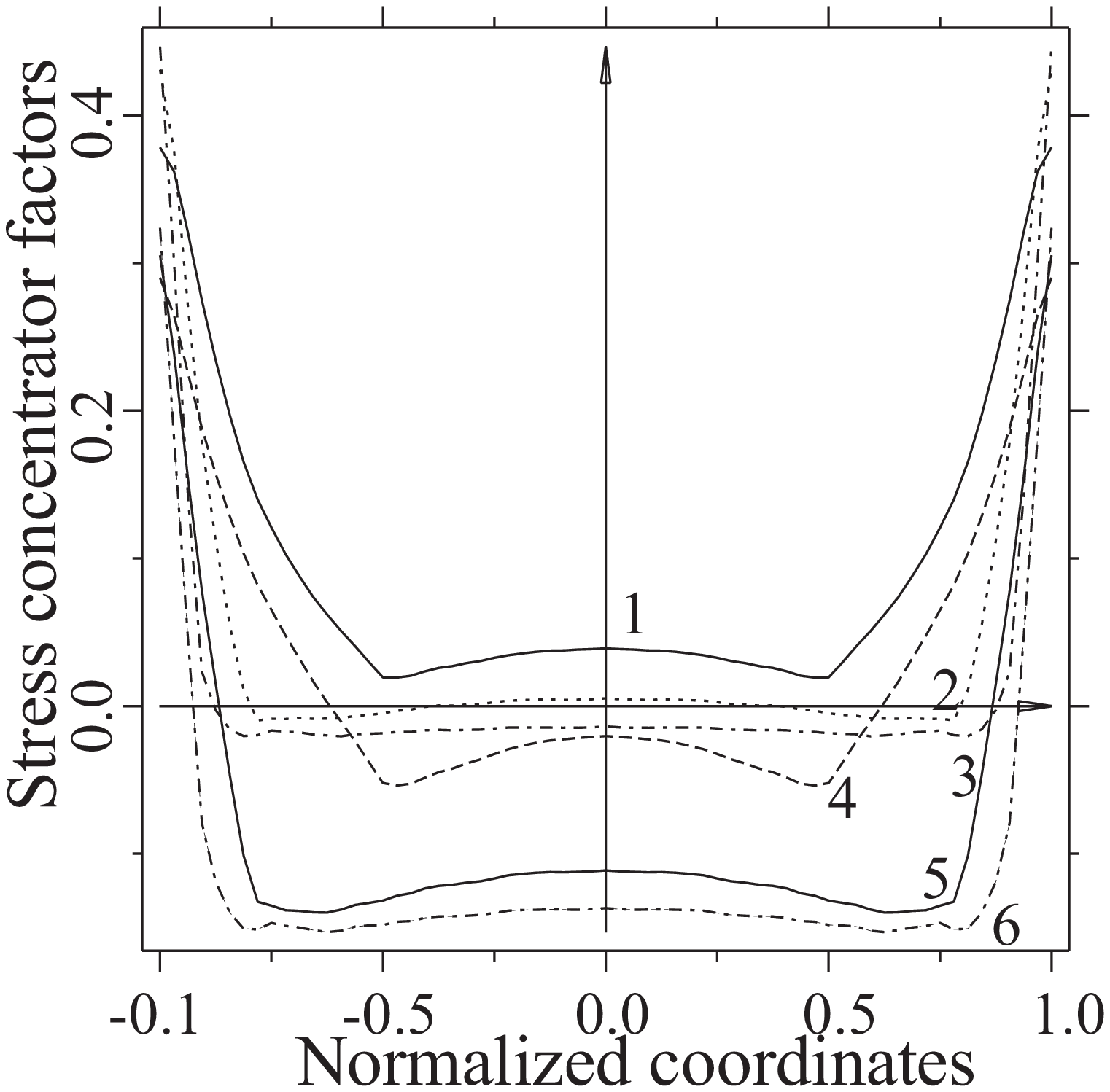,width=\linewidth}\\
\vspace{0.0cm}
\tenrm {\sc Fig. 9}:
%Stress concentrator factors normalized coordinate
$B^{*[k]}_{1122}(x_1)$ vs
$x_1/a$ for $k=0$: 1 ($h/a^c=1.0$), 2 ($h/a^c=0.25$), 3 ($h/a^c=0.1$) and for $k=12$:   4 ($h/a^c=1.0$), 5 ($h/a^c=0.25$), 6 ($h/a^c=0.1$).
%\hspace{1cm}
\end{minipage}\hfill
\vspace{0.0mm}
\begin{minipage}[b]{.42\linewidth}
\centering\epsfig{figure=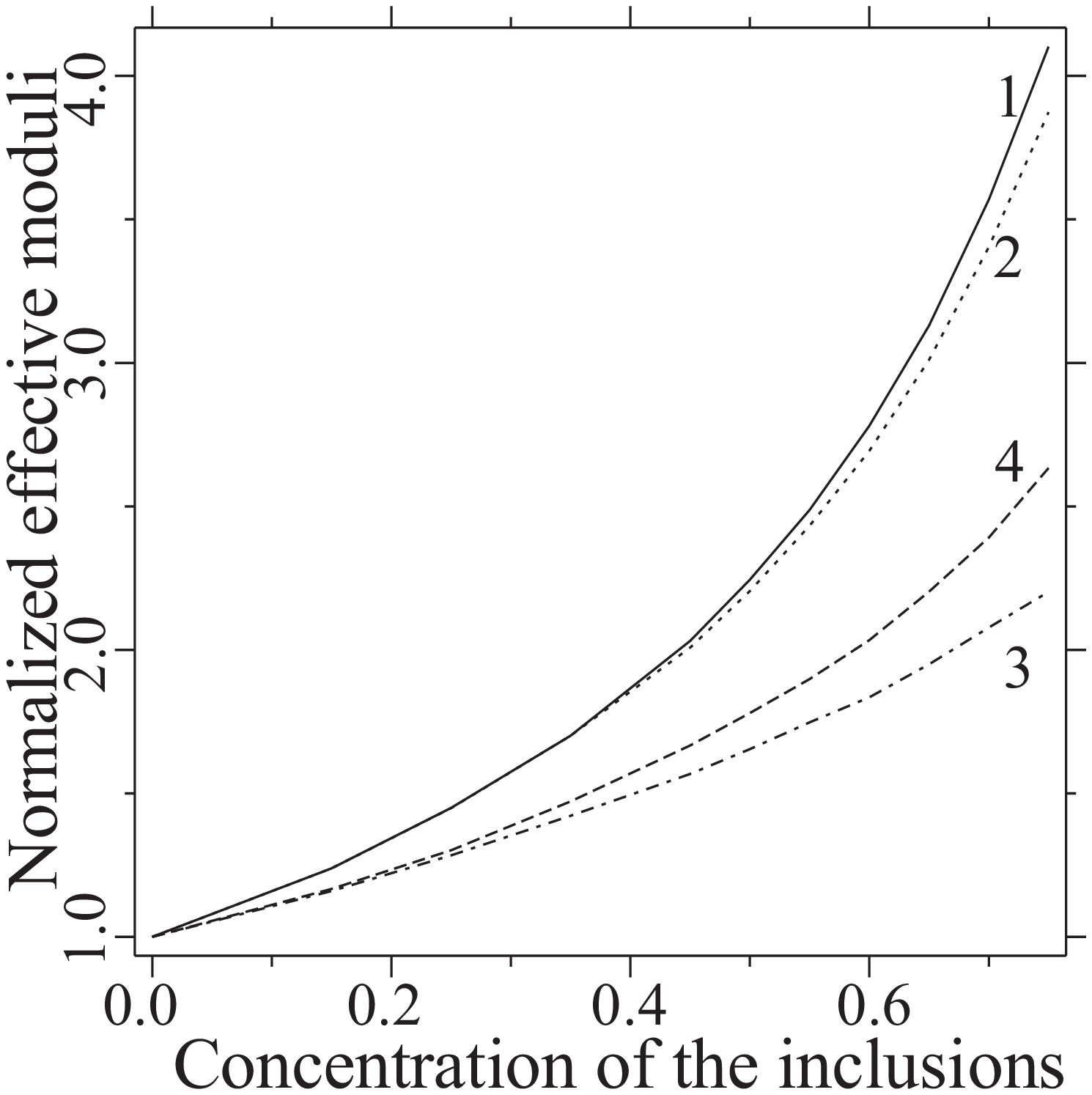,width=\linewidth}\\
\vspace{0.0cm}
\tenrm {\sc Fig. 10}: Normalized effective moduli $\mu^*/\mu^{(0)}$ (1,2) and $k^*/k^{(0)}$ (3,4)
estimated by the use of both the classical approach (2,4) and new one (1,3)
\end{minipage}

\medskip
\noindent{\bf 7. Conclusion}
\medskip

We have proposed the new background of micromechanics based on the new general integral equation (4.1) which does not use the central concept of classical micromechanics such as effective field hypothesis {\bf H1}. The eventual abandonment from hypothesis {\bf H1}
has made a rejection of the satellite hypothesis {\bf H3} possible. If statistical averages of stresses in the heterogeneities can be considered as homogeneous ones then the new approach is degenerated into the classical approach (4.19)-(4.21). However such an assumption is approximately appropriate only for statistically homogeneous fields of homogeneous ellipsoidal inhomogeneities subjected to homogeneous boundary conditions and fulfilled at the conditions of quasi-crystallite approximation 3. If any of the indicated conditions is broken then an appearing inhomogeneity of stress fields in the inclusions lead to one of two possible sources of inhomogeneities of the effective field which, in turn, generates an additional inhomogeneity of stress fields inside inclusions and so on. For
%\noindent
example, if all above-listed conditions are satisfied but the closing hypothesis {\bf H2b} is replaced by the hypothesis {\bf H2a} taking the binary interaction of inclusions into account then the first sort of inhomogeneity of the effective field $\lle\overline{\bfsi}\rle(\bfx)$ is generated by the binary interaction of inclusions [see the item $\bfcL_q^{\sigma}(\lle\overline{\bfsi}|;v_i,\bfx_i\rle_{q})(\bfx)$ in Eq. (4.6)] even if this interaction is approximately estimated through the matrix $\bfZ$ as in Eq. (2.18). However, even in the framework of hypothesis {\bf H2b}, the replacement of ellipsoidal homogeneous inclusions by either the nonellipsoidal homogeneous ones or inhomogeneous (e.g. coated) ellipsoidal inclusions with necessity leads to the second sort of effective field inhomogeneity produced by the fundamentally new renormalizing item $\lle\bfcL^{\sigma}(\bfet)\rle(\bfx)$ (4.1).
This new renormalizing item is directly dependent [in opposite to the classical Eq. (2.2)] on inhomogeneity of stress fields inside the inclusions that has lead to detection of fundamentally new effects in micromechanics such as dependence of stress concentrator factors estimated (see Figs. 7 and 8) on both the RDF and size of the excluded volume even in the framework of hypotheses {\bf H2b} and {\bf H3}.

The modeling and simulation of random nano- and microstructures are becoming
more and more ambitious due to the advances in modern computer software
and hardware that is stimulated by a real challenge of modern material science and technology. The researches can forget about restrictions of analytical solutions (such as, e.g., Eshelby tensor and hypothesis {\bf H1}) and use the numerical solutions
which they need. It is expected to get all the more differences between the old and new approaches than inhomogeneity of the stress concentrator factors $\bfB_i(\bfx)$ ($\bfx\in v_i$) would be larger. So, for the square inclusion with the smoothed vertexes and the finite cylindrical fiber the components $\bfB_i(\bfx)$ can vary by factors of four and ten (compare with Fig. 2), respectively (see Subsections 4.2.4 and 18.3.2, respectively,  in Ref. [6]).
%Buryachenko, 2007).
Another source of stress inhomogeneity inside the inclusions is a continuous variation of their mechanical properties such as in either cylindrically or spherically anisotropic particles (see, e.g., [48]). %Chen, 1993).
However, probably the most
often investigated reason of such a stress inhomogeneity is an imperfect interphase
(including sliding, debonding, cohesive phenomena, see for references, e.g., [6]).  %Buryachenko, 2007).
These interphase may represent weak interfacial layer due to imperfect bonding
between the two phases and inter-diffusion and/or chemical interaction zones
(with properties varying through the thickness and/or along the surface) at the
interphase between the two phases. The thickness of interphase investigated usually ranges from $h/a^c=0.01$ for the conventional composites to $h/a^c=2$ for nanocomposites.
The significance of interphase effects becomes important in nanocomposites due to their high surface-to-volume ratios. An alternative approach taking into account interfacial effect is based on the concept of surface stress and surface tension (see, e.g.,
 [49], [50]). % Sharma and Ganti, 2004; Chen {\it et al.}, 2007).
To the author's knowledge, in tens of publications dedicated to the influence of interphase on effective properties,  the methods usually based on the hypothesis {\bf H1} (such as, e.g., the Mori-Tanaka scheme and MEF) are exploited. Now all these estimations can be improved in the framework of the new approach as we did it in Figs. 2-10 [compare the results obtained for the initial $\bfB^{*[0]}(\bfx)$ and 12th $\bfB^{*[12]}(\bfx)$ iterations].

Other possible directions of successful applications of the proposed approach are three classes of problems where inhomogeneities of stress distributions in the inclusions are generated by the nonlocal effects even for homogeneous ellipsoidal inclusions. The first two classes of these problems are
described by both the
special features of  applied loading
(statistically homogeneous media subjected to inhomogeneous boundary conditions) and the special features of microstructure
(FGMs, clustered materials, bounded media, contact of microinhomogeneous media, macro-heterogeneity insde microinhomogeneous medium, see for details and references [6]). In both cases, the known methods are based on the general integral equation (2.2) for the statistically inhomogeneous media when  $\lle\bfet\rle(\bfy)\not \equiv $const.
%(1990)
(see for details and references [6]). %Buryachenko, 2007).
%    && ; particular cases of Eq. (2.2) were also widely used (either explictly
%    && or implicitly) by other authors: (Drugan and Willis, 1996;
%    &&  Drugan, 2000, 2003; Luciano and Willis, 2003, 2004; Sharif-Khodaei and Zeman, 2008).
However, Eq. (2.2) is just an approximation obtained from the exact Eq. (4.1) at the assumption (3.23I). Using of more general Eq. (4.1) instead of the approximative Eq. (2.2)
opens up great opportunities for detection of new effects in nonlocal micromechanics. The mentioned problems imply an estimation of nonlocal effective properties for composites through their constituents exhibit local constitutive properties. A new inverse problem is initiated by investigation of nanocomposites and formulated as an estimation of local effective properties through the nonlocal mechanical properties of constituents.
This problem was solved by Buryachenko [6]
%(2007)
(see Section 18.2) in the framework of hypothesis {\bf H1}. However, it is well known in the context of micropolar elasticity that the strains are non-uniform even for the homogeneous elastic properties of the ellipsoidal inclusion subjected to the homogeneous remote loading. This sort of inhomogeneity is an encouragement for generalization of Eq. (4.1) to the composites which constituents are described by the nonlocal constituent laws. A subsequent step is the adoption of the new approach proposed in this paper for analysis of the generalized Eq. (4.1). However,
more detailed consideration of nonlocal effects mentioned is beyond the scope of the
current study and will be analyzed in other publications.

% Material properties of Fused Silica in $\mu$MKS units:
% Young's Modulus 74,52*$10^3$   Poisson's Ratoio 0.17

% Glass fiber (E-glass) E(f)Z77 GPa, n(f)Z0.2, r(f)Z6 mm
% Epoxy resin 556/917 E(m)Z3.11 GPa n(m)Z0.34
% Interphase thickness t(i)Z0.3–0.7 mm

\medskip
\noindent{\bf Acknowledgments:}
\medskip

This work was partially supported by  the Visiting Professor Program of the University of Cagliari and the  Eppley Foundation for Research.

\medskip
\noindent{\bf References}
\medskip
{\baselineskip=9pt
\parskip=1pt
\lrm

\hangindent=0.4cm\hangafter=1\noindent
\ 1. Willis, J. R.:
Variational and related methods for the overall properties
of composites. {\tenitr Advances in Applied Mechanics},
{\tenbf 21}, 1--78 (1981)

\hangindent=0.4cm\hangafter=1\noindent
\ 2. {\tensc Mura, T.}:   {\tenitr Micromechanics of
Defects in Solids}. Martinus Nijhoff, Dordrecht (1987)

\hangindent=0.4cm\hangafter=1\noindent
\ 3. {Nemat-Nasser, S.}, {Hori, M.}:  {\tenitr Micromechanics: Overall Properties of Heterogeneous Materials}. Elsevier, North-Holland (1993)

\hangindent=0.4cm\hangafter=1\noindent
\ 4. Torquato, S.: {\tenitr Random Heterogeneous Materials:
Microstructure and Macroscopic Properties.}
Springer-Verlag (2002)

\hangindent=0.4cm\hangafter=1\noindent
\ 5. Milton, G. W.:  {\tenitr The Theory of Composites.} Appl. Comput. Math., v. {\tenbf 6}, Cambridge University Press (2002)

\hangindent=0.4cm\hangafter=1\noindent
\ 6. Buryachenko, V.A.: {\tenitr Micromechanics of Heterogeneous Materials}. Springer, NY (2007)

\hangindent=0.4cm\hangafter=1\noindent
\ 7. Kanaun, K. K., Levin, V. M.:  {\tenitr Self-Consistent Methods for Composites}.
Vol. 1, 2. Springer, Dordrecht (2008)

\hangindent=0.4cm\hangafter=1\noindent
\ 8. Kr\"oner, E.:
 Berechnung der elastischen Konstanten des Vielkristalls
aus den Konstanstanten des Einkristalls.
{\tenitr Z. Physik.} {\tenbf 151}, 504--518 (1958)

\hangindent=0.4cm\hangafter=1\noindent
\ 9. {Hill, R.}:  A self-consistent mechanics of composite
materials. {\tenitr   J. Mech. Phys. Solids} {\tenbf 13}, 212--222 (1965)

\hangindent=0.4cm\hangafter=1\noindent
10. {Mori, T.}, {Tanaka, K.}: Average stress in matrix and average elastic energy of materials with misfitting inclusions. {\tenitr  Acta Metall}. {\tenbf 21}, 571--574 (1973)

\hangindent=0.4cm\hangafter=1\noindent
11. {Benveniste, Y.}:   A new approach to application of
Mori-Tanaka's theory in composite materials. {\tenitr  Mech. Mater.}
{\tenbf 6}, 147--157 (1987)

\hangindent=0.4cm\hangafter=1\noindent
12. Morse,~P. M., Feshbach, ~H.:     {\tenitr      Methods of Theoretical Physics.}
Parts I and II. McGraw-Hill, Maidenhead (1953)

\hangindent=0.4cm\hangafter=1\noindent
13. Mossotti, O. F.:  Discussione analitica sul'influenza che l'azione di un mezzo dielettrico ha sulla distribuzione dell'electricit\'a alla superficie di pi\'u corpi elettrici disseminati in eso.    {\tenitr Mem Mat Fis della Soc Ital di Sci in Modena,} {\tenbf      24}, 49--74 (1850)

\hangindent=0.4cm\hangafter=1\noindent
14. Markov,~K. Z.:  Justification of an effective field method in elasto-statics of heterogeneous solids. {\tenitr J Mech Phys Solids}, {\tenbf 49}, 2621--2634  (2001)

\hangindent=0.4cm\hangafter=1\noindent
15. Scaife,~B. K. P.:      {\tenitr Principle of Dielectrics.}
Oxford University Press, Oxford, UK (1989)

\hangindent=0.4cm\hangafter=1\noindent
16. Lax, M.:     Multiple scattering of waves II. The effective
fields dense systems. {\tenitr  Phys. Rev.} {\tenbf  85}, 621--629 (1952)

\hangindent=0.4cm\hangafter=1\noindent
17. Foldy,~L.L.:   The multiple scattering
of waves. I. General theory of isotropic scattering by randomly
distributed scatters.  {\tenitr  Phys Rev,} {\tenbf      67}, 107--117 (1945)

\hangindent=0.4cm\hangafter=1\noindent
18. Chaban,~I.A.:   Self-consistent field approach to calculation of
the effective parameters of microinhomogeneous media.
  {\tenitr    Akust  Zhurn,} {\tenbf    10}, 351--358 (1965)
(In Russian. Engl  Transl.      {\tenitr      Soviet Physics-Acoustics,} {\tenbf
   10}, 298--302 (1965))

\hangindent=0.4cm\hangafter=1\noindent
19. Walpole, L.J.:   On the bounds for the overall elastic moduli of inhomogeneous system. I, II.     {\tenitr       J Mech Phys Solids,} {\tenbf   14}, 151--162, 289--301 (1966)

\hangindent=0.4cm\hangafter=1\noindent
20. Horii, H., Nemat-Nasser, S.: Elastic field of interacting inhomogeneities.
{\tenitr Int J Solids Struct}, {\tenbf 21}, 731-–745 (1985)

\hangindent=0.4cm\hangafter=1\noindent
21. Buryachenko,~V.A.,  Rammerstorfer,~F.G.:
 On the thermostatics of composites with coated inclusions.
   {\tenitr    Int  J  Solids  Struct,} {\tenbf        37}, 3177--3200 (2000)

\hangindent=0.4cm\hangafter=1\noindent
22. Willis, J.R.:
Bounds and self-consistent estimates for the overall properties
of anisotropic composites.  {\tenitr J. Mech. Phys. Solids}, {\tenbf 25}, 185--203 (1977)

\hangindent=0.4cm\hangafter=1\noindent
23. Eshelby,~J.D.:   The determination of the elastic field of an ellipsoidal inclusion, and related problems.
{\tenitr Proc  Roy Soc  Lond,} {\tenbf   A241}, 376--396 (1957)

\hangindent=0.4cm\hangafter=1\noindent
24. Buryachenko, V.A. (referred to as (I) in the text):
On the thermo-elastostatics of heterogeneous materials. I. General integral equation.
{\tenitr Acta Mech}. (2009)
(Submitted)

\hangindent=0.4cm\hangafter=1\noindent
25. Willis,~J.R.,  Acton,~J. R.:   The overall elastic moduli of a dilute suspension of spheres.     {\tenitr  Q  J  Mechan  Appl  Math,} {\tenbf 29}, 163--177 (1976)

\hangindent=0.4cm\hangafter=1\noindent
26. Dvorak, G.J., Benveniste, Y.:
On transformation strains and uniform fields in multiphase elastic media.
{\tenitr    Proc   Roy Soc  Lond,} {\tenbf    A437}, 291--310 (1992)

\hangindent=0.4cm\hangafter=1\noindent
27. Nogales, S., B\"ohm, H.J.:
Modeling of the thermal conductivity and thermomechanical behavior of diamond reinforced composites. {\tenitr Int. J. Engng. Sci.} {\tenbf 46}, 606--619 (2008)

\hangindent=0.4cm\hangafter=1\noindent
28. {Khoroshun, L.P.}:
Random functions theory in problems on the macroscopic
characteristics of microinhomogeneous media.
{\tenitr Priklad. Mekh. }, {\tenbf 14}(2),
3--17 (1978). (In  Russian. Engl. Transl. {\tenitr Soviet Appl. Mech.}
{\tenbf 14}, 113--124 (1978))

\hangindent=0.4cm\hangafter=1\noindent
29. Buryachenko,~V.A.,     Parton,~V.Z.:
One-particle approximation  of the effective field method in the
statics of composites.   {\tenitr    Mekh  Kompoz Mater,}    (3), 420--425 (1990)
(In Russian. Engl Transl.    {\tenitr   Mech Compos Mater,} {\tenbf         26}(3), 304--309
(1990))

\hangindent=0.4cm\hangafter=1\noindent
30. Kanaun, S.K.:  Elastic medium with random fields of inhomogeneities. In:
Kunin, I. A. {\tenitr Elastic Media with Microstructure}. Springer–Verlag, Berlin, {\tenbf 2}, 165-–228 (1983)

\hangindent=0.4cm\hangafter=1\noindent
31.  Khoroshun, L.P.:
Prognosis of thermoelastic properties of materials reinforced by unidirectional discrete fibers. {\tenitr Priklad. Mekh. }, {\tenbf 10}(12),
23--30 (1974). (In  Russian. Engl. Transl. {\tenitr Soviet Appl. Mech.}
{\tenbf 10}, (1974)).

\hangindent=0.4cm\hangafter=1\noindent
32. Ponte Casta\~neda, P.,  Willis, J.R.:
The effect of spatial distribution on the
effective behavior of composite materials and cracked media.
{\tenitr J. Mech. Phys. Solids}, {\tenbf 43}, 1919--1951 (1995)

\hangindent=0.4cm\hangafter=1\noindent
33. Markov,~K.Z.:   Elementary micromechanics of heterogeneous media.
    In: Markov~K,  Preziosi~L (eds),     {\tenitr Heterogeneous Media. Micromechanics, Modeling, Methods, and Simulations.}  Birkh\"auser, Boston,   1--162 (1999)

\hangindent=0.4cm\hangafter=1\noindent
34. Delves, L.M., Mohamed,~J.L.:
        {\tenitr   Computational Methods for Integral Equations.}   Cambridge University Press, Cambridge, UK (1985)

\hangindent=0.4cm\hangafter=1\noindent
35. Buryachenko,~V.A.,  Tandon,~G.P.:  Estimation of effective elastic properties of random structure composites for arbitrary inclusion shape and anisotropy of components using finite element  analysis.   {\tenitr  Int J Multiscale Comput Engng,} {\tenbf 2}, 29--45 (2004)

\hangindent=0.4cm \hangafter=1\noindent
36.  Chen, H.S., Acrivos, A.:
The effective elastic moduli of composite materials containing spherical
inclusions at non-dilute concentrations.
{\tenitr Int. J. Solids Structures}, {\tenbf 14}, 349-364(1978)

\hangindent=0.4cm\hangafter=1\noindent
37. Perlin,~P.I.:   Application of the regular representation
of singular integrals
to the solution of the second fundamental problem of the theory of elasticity.
   {\tenitr    Prikl Metem Mekhan,} {\tenbf    40}, 366--371 (1976) (In Russian. Engl Transl.
      {\tenitr  J Appl Math Mech,} {\tenbf    40}, 342--347 (1976))

\hangindent=0.4cm\hangafter=1\noindent
38. Mikhlin,~S.G., Pro\"ossdorf,~S.:  (1980)     {\tenitr    Singular Integral Operators.}
Springer-Verlag, Berlin, New York (1980)

\hangindent=0.4cm\hangafter=1\noindent
39. Varga,~R.S.: {\tenitr     Matrix Iterative Analysis.}
Springer, Berlin (2000)

\hangindent=0.4cm\hangafter=1\noindent
40. Hansen,~J.P., McDonald,~I.R.:    {\tenitr        Theory of Simple Liquids.}
Academic Press, New York (1986)

\hangindent=0.4cm\hangafter=1\noindent
41. Torquato,~S., Lado,~F.:   Improved bounds on
the effective elastic moduli of random arrays of cylinders.
         {\tenitr  J Appl Mech,} {\tenbf    59}, 1--6 (1992)

\hangindent=0.4cm\hangafter=1\noindent
42. Hirai, T., Sasaki, M., Niino, M.:  CVD in-situ ceramic composites. {\tenitr J. Mater. Sci. Soc. Japan}
{\tenbf 36}, 1205-–1211 (1987)

\hangindent=0.4cm\hangafter=1\noindent
43. Theocaris, P.S.:  {\tenitr The Concept of Mesophase in Composites}. Berlin, Springer
(1987)

\hangindent=0.4cm\hangafter=1\noindent
44. Jayaraman, K., Reifsnider, K.L.:  Residual stresses in a composite
with continuously varying Young's modulus in the fiber/matrix interphase. {\tenitr J. Comp. Mater.} {\tenbf 26}, 770-–791 (1992)

\hangindent=0.4cm\hangafter=1\noindent
45. Wang, W., Jasiuk, I.:  Effective elastic constants of particulate composites with inhomogeneous
interphases. {\tenitr J. Comp. Mater.}, {\tenbf 32}, 1391-–1424 (1998)

\hangindent=0.4cm\hangafter=1\noindent
46. Weng, G.J.: Effective bulk moduli of two functionally
graded composites. {\tenitr Acta Mechanica}, {\tenbf 166}, 57–-67 (2003)

\hangindent=0.4cm\hangafter=1\noindent
47. You, L.H., You, X.Y., Zheng, Z.Y.:
Thermomechanical analysis of elastic–plastic fibrous
composites comprising an inhomogeneous interphase
{\tenitr Computational Materials Science}, {\tenbf 36},  440–-450 (2006)

\hangindent=0.4cm\hangafter=1\noindent
48. Chen, T.:  Thermoelastic properties and conductivity of composites reinforced
by spherically anisotropic inclusions. {\tenitr Mechan of Mater}, {\tenbf 14}, 257-–268
(1993)

\hangindent=0.4cm\hangafter=1\noindent
49. Sharma, P., Ganti, S.:  Size-dependent Eshelby's tensor for embedded nanoinclusions
incorporating surface/interface energies. {\tenitr  J Appl Mech}, {\tenbf 71}, 663-–671
(2004)

\hangindent=0.4cm\hangafter=1\noindent
50. Chen, T., Dvorak, G.J., Yu, C.C.:  Size-dependent elastic properties of unidirectional nano-composites with interface stresses.
{\tenitr Acta Mechan}, {\tenbf 188}, 39-–54 (2007)

\hangindent=0.4cm\hangafter=1\noindent
% &&  Sharif-Khodaei, Z, Zeman, J. (2008) Microstructure-based modeling of elastic functionally graded materials:
% &&  one dimensional case. {tenit J. Mechanics of Materials and Structures}, {\tenbf 3}, 1773--1796

%   && \hangindent=0.4cm\hangafter=1\noindent
%   Willis, J. R. (1982)   Elasticity theory of composites. {\tenit Mechanics of
%   Solids. The Rodney Hill 60th Anniversary Volume}
%   (eds. H. G. Hopkins  and  M. J. Sewell), pp. 653--686,  Pergamon Press, Oxford

%   && \hangindent=0.4cm\hangafter=1\noindent
%   && {Willis, J. R.}  (1983)  The overall elastic response of composite
%   && materials.  {\tenit ASME J. Appl. Mech.} {\tenbf  50}, 1202--1209

%   && \hangindent=0.4cm\hangafter=1\noindent
%   && Willis, J. (1984) Some remarks on the application of the QCA to the determination
%   && of the overall elastic response of a matrix/ inclusion composite. {\tenit J Math
%   && Phys}, {\tenbf 25}, 2116-–2120

}
\end{document}